\documentclass{aastex}

\pdfoutput=1
\usepackage{mathptmx}
\usepackage{natbib}
\shorttitle{Dust cores around bubbles N14, N22, and N74}

\newcommand{\nh}{N$_2$H$^+$}
\newcommand{\hco}{HCO$^+$}
\newcommand{\hii}{\ion{H}{2}~}

\begin{document}

\title{Investigation of Molecular Cloud Structure around Infrared Bubbles: CARMA
Observations of N14, N22, and N74}
\author{Reid A. Sherman}
\affil{Department of Astronomy \& Astrophysics, University of Chicago, 5640 S. Ellis Ave.,
Chicago, IL 60637}
\email{ras@oddjob.uchicago.edu}

\begin{abstract}
We present CARMA observations in 3.3 mm continuum and several molecular lines of the surroundings of N14, N22, and N74, three infrared bubbles from the GLIMPSE catalog. We have discovered 28 compact continuum sources and confirmed their associations with the bubbles using velocity information from \hco ~and HCN. We have also mapped small-scale structures of \nh ~emission in the vicinity of the bubbles. By combining our data with survey data from GLIMPSE, MIPSGAL, BGPS, and MAGPIS, we establish about half of our continuum sources as star-forming cores. We also use survey data with the velocity information from our molecular line observations to describe the morphology of the bubbles and the nature of the fragmentation. We conclude from the properties of the continuum sources that N74 likely is at the near kinematic distance, which was previously unconfirmed. We also present tentative evidence of molecular clouds being more fragmented on bubble rims compared to dark clouds, suggesting that triggered star formation may occur, though our findings do not conform to a classic collect-and-collapse model.

\end{abstract}

\keywords{ISM: bubbles---ISM: \hii regions---ISM: kinematics and dynamics---stars: formation}

\section{INTRODUCTION}
\label{cha:intro}

Despite their relatively small numbers, massive stars play a crucial role in the chemistry and physics of the interstellar medium. They are the dominant source of ionizing photons and heavy elements, and the shock waves created by associated \hii regions, stellar winds, and supernovae provide momentum that drives the dynamical evolution of the gas. As they only form in tight clusters within massive molecular clouds, feedback processes may play an important role in both star formation and cloud evolution. Because of their scarcity and the density of their environments, massive star formation is challenging to study observationally. Massive star-forming regions are statistically more distant, and high dust extinctions at short wavelengths and source confusion at longer wavelengths further complicate matters. Hence, the stages of cluster formation are still not well understood.

One important process in massive star formation is the creation of compact \hii regions and the interaction of the \hii region with the surrounding cloud. An ionization shock expanding into a dense molecular cloud can increase the density and pressure in a shell around the \hii region, possibly resulting in fragmentation into compact star-forming cores \citep{EL77}. This model, called collect-and-collapse, is a well-studied mechanism by which the presence of a massive star can trigger new formation in the local neighborhood.

Analytical models and computer simulations have tested many aspects of the collect-and-collapse process. A threshold mass of the central source to trigger successive generations of \hii regions has been proposed \citep{HI06b}, as well as a distribution of the masses of self-gravitating clumps \citep{DBW07}. \citet{HI06a} argued that the fragmentation should depend strongly on the ambient density but only weakly on the mass of the source driving the shock. Other recent simulations have cast doubt on the whole collect-and-collapse model, saying that protostars in a shell surrounding an \hii region could plausibly be pre-existing density fluctuations in the ambient cloud left visible as the \hii region blows out the low density areas \citep{Walch11}.

GLIMPSE \citep{GLIMPSE} and MIPSGAL \citep{MIPSGAL}, two infrared galactic plane surveys from the Spitzer Space Telescope, form an invaluable database for studying whether triggered star formation is common in our Galaxy. By covering the inner third of the galactic plane in a stripe 2 degrees wide, they provide high-resolution infrared data across numerous bands of many high-mass star forming regions, which tend to be confined to low galactic latitudes. One of the striking features from these surveys was a large population of ``bubbles'' (\citealt{Ch06}, hereafter Ch06), rings of mid-infrared emission, strongest in the 8-\micron ~band. Many of these were found by MIPSGAL to contain extended 24-\micron ~emission confined to the inside of the bubble, which is thought to come from warm dust evaporated into the diffuse bubble interior \citep{Everett10}. The close correlation of bubbles with \hii regions (Ch06; \citealt{Deharveng10}, hereafter D10) has led to the generally accepted view that the bubbles are infrared signatures of heated regions around young massive stars and star clusters with the 8-$\micron$ ring tracing the photo-dissociation region (PDR), which directly borders where a shell would be expected to form in the collect-and-collapse process. With the added information of the 24-\micron data and more visual inspections, the catalog of known bubbles has increased to more than 5000 \citep{Simpson12}, so understanding how these objects affect their surroundings has big implications for Galactic star formation. In this paper, we will use the term ``bubble'' to refer to a region, whether or not it appears in the Ch06 catalog, that has extended 24 $\micron$ emission surrounded by at least a partial ring of 8 $\micron$ emission.

Many people have claimed to have found observational evidence for triggered star formation. Even before the Spitzer surveys, near-infrared observations of \hii regions and prominent pillars (e.g.~\citealp{Sugitani02}) attempted to find a statistical segregation of stellar ages, suggesting sequential triggering by a passing shock front. This method has led to mixed, contradictory results (e.g.~\citealt{Ind07,Guarcello10}). \citet{Elm11} points out that gravitational instabilities due to the dense shell could drag some newly formed stars along with it, causing any age segregation to be difficult to distinguish. Recently there have been many papers that claim to find triggered protostars on the edge of bubbles in multi-wavelength datasets (e.g.~\citealp{Deharveng08}), some including the Spitzer surveys (e.g.~\citealt{Pom09,Petriella10}) and new Herschel observations \citep{Zavagno10b}.

There are shortcomings of these techniques, however. YSOs found in the neighborhoods of bubbles are suggestive, but are not clearly indicative, of triggered star formation. The YSOs may be objects in the line of sight or might have formed even without the trigger of the nearby \hii region. \citeauthor{BW10} (2010, hereafter BW10) found that although many bubbles had YSOs nearby, only two of the 43 bubbles they analyzed had a clear statistical overdensity of YSOs. To add more weight to tie ongoing star formation to a specific trigger requires more spectral information to map the velocity structure of the molecular cloud and fragments, and sub-millimeter and millimeter observations to track earlier stages of cloud fragmentation and get a more complete picture of the evolution of a shell through the fragmentation and star formation process.

There is some dispute about the general morphology of GLIMPSE bubbles as well, which could affect the likelihood of triggered star formation. \citet{Watson08} show that the 8-$\micron$ profile of N49 is fit very well with a three-dimensional shell model, and conclude that many bubbles are spheroidal objects with bright rims at the edge because of the greater column density. BW10, on the other hand, analyze bubbles by their CO emission and find no evidence of a three-dimensional shell. Their CO profiles are not well-fit at all by a shell model and they conclude that the bubbles are predominantly open ring-shaped objects embedded in sheet-like molecular clouds.

The Bolocam Galactic Plane Survey (BGPS; \citealp{BGPS}), a 1.1-mm continuum survey of the northern galactic plane, and the Galactic Ring Survey (GRS; \citealp{GRS}), a survey of $^{13}$CO emission in the northern galactic plane, add valuable information about the surroundings of the GLIMPSE bubbles. Their mapping of dust and molecular gas show bright spots at the edges of a number of bubbles, indicating promising sites of fragmentation and triggered star formation, although their low angular resolution ($>$ 30$\arcsec$) precludes detailed analysis of the structure of bubble shells.

Millimeter-wave interferometry is necessary to investigate the early stages of star formation in these shock-impacted clouds. Interferometers can achieve much higher spatial resolution and filter out the extended structure to be especially sensitive to sub-parsec to parsec scale fragments in the clouds. Millimeter wavelengths are useful because while far from the peak of the spectral energy distribution for clumps at the temperatures of interest ($\sim$10-100K), the dust will often still be bright enough to detect in thermal continuum emission, and there are an abundance of molecular lines of interest. These lines can be used to measure the velocity structure of the clumps and confirm the association of embedded cores with the bubbles, as well as trace the dynamical and chemical processes in the early stages of triggered star formation. \citet{Fukuda02} mapped a region of M16 in isotopes of CO and found significant fragmentation in the pillars and concluded that at least a couple of the fragments were in the process of forming stars. Similar observations of \hii regions in earlier stages of formation (i.e. GLIMPSE bubbles) have been lacking, however, and recent advances in instrumentation have opened up many new lines of investigation.

As well as at the rims of bubbles, infrared dark clouds (IRDCs) have been proposed as sites of likely massive star formation \citep{Chambers09}. They are frequently bright in millimeter wavelengths (e.g. \citealp{Beuther05}), and GLIMPSE Extended Green Objects (EGOs, \citealt{EGOs}), which seem to be associated with early stages of massive star formation \citep{Cyg11}, are often located in IRDCs. \citet{IRDCs} catalogued thousands of IRDCs, and, unsurprisingly, there is correlation in the distribution of IRDCs and bubbles, with many bubbles having IRDCs in the vicinity. One open question in triggered star formation is whether the triggering process results in a different stellar initial mass function than would happen without the trigger. By observing bubble rims and nearby IRDCs, we can begin to see if there are any systematic differences in the fragmentation of molecular clouds.

The Combined Array for Research in Millimeter-wave Astronomy (CARMA; \citealp{CARMA}) is well-suited for such observations. With eight independently customizable bands, many lines can be observed simultaneously while still giving significant bandwidth to continuum to detect dense dust cores. This is a significant advance from just two years ago, when there were only three bands and the velocity channels were  $\sim$6 times wider at equivalent bandwidth settings. The 3-mm observing window contains a plethora of lines that provide good diagnostics for star formation and cloud dynamics. While GLIMPSE bubbles are often very large compared to the 3-mm single-point field of view on CARMA ($\sim$1$\arcmin$), CARMA's well-developed mosaicing procedures allow for covering large areas with even noise across the field.

In this paper we use exploratory CARMA observations of the fine scale structure around three infrared bubbles to address three outstanding questions. The first is what the velocity structure of the molecular line emission can tell us about the three-dimensional structure of the bubbles. The second is what the dust structures we see tell us about star formation on the rims of bubbles. The third is whether bubble shells are fragmenting to a greater extent than uncompressed clouds, as the star formation at the rims of bubbles can only truly be called ``triggered'' if the stars would not have formed without the effect of the expanding \hii region.

While our sample of objects is only three, our observations reveal an abundance of structure on bubble rims at scales significantly smaller than had previously been probed. The resolution of our maps is comparable to Spitzer's survey maps, significantly better than previous millimeter-wave observations, so for the first time we can more accurately correlate molecular cloud structures with structures seen in the infrared.

\section{OBSERVATIONS}
\label{cha:obs}

\subsection{Source Selection}

In this paper, we describe our observations of three bubbles: N14, N22, and N74, as listed in Table \ref{table-sources}. (The naming scheme is from the Ch06 catalog.) N14 was picked as a test subject because of its relatively near kinematic distance (3.7 kpc; Ch06), the large millimeter flux seen in the BGPS, and the presence of ammonia emission (Cyganowski et al., in prep). N22 and N74 were picked as additional sources as they satisfied a number of conditions:

\begin{itemize}
\item 
Kinematic distances of less than 5 kpc, so that the observations would probe size scales significantly smaller than seen before.
\item
Strong millimeter flux seen in the BGPS, which also determined the extent of the area we chose to map with CARMA.
\item
Presence of bright $^{13}$CO emission seen in the GRS.
\item
Angular sizes small enough to cover with a CARMA mosaic of $\sim$60 pointings or fewer, allowing reasonable signal-to-noise ratio in $\sim$20 hours of observation per object.
\item
Observations by BW10, who derived properties of the \hii regions in and molecular clouds surrounding 43 northern-hemisphere bubbles.
\end{itemize}

BW10 measured the total free-free emission flux from the MAGPIS data and used it to estimate the strengths of the sources driving the bubbles in their catalog, which included N14, N22, and N74. They put this in terms of the number of O9.5 stars that would be required to explain the ionizing flux. This number will be a lower limit, as some UV flux may be absorbed by dust or escape from the \hii region and not result in free-free emission. For N14, they calculated the necessary number of O9.5 stars is 6.4, while for N22 it is 9.8 and for N74 0.03. For comparison, the number for the Orion nebula would be 20. For the 40 \hii regions they analyzed, the median was 1.0 O9.5 star and only four were calculated to be driven by a more powerful driving source than the Orion nebula.

\begin{deluxetable}{ccccccc}
\footnotesize
\tablecaption{Observed Bubbles
\label{table-sources}
}
\tablewidth{0pt}
\tabletypesize{\scriptsize}
\tablehead{
Source\tablenotemark{a} & $\alpha$(J2000)  & $\delta$(J2000) & $l$ &
$b$ & $v$$_{\textrm{CO}}$\tablenotemark{b} & Distance\tablenotemark{c} \\
  &   &  &  &  & (km s$^{-1}$) & (kpc) 
}
\startdata
N14 & 18$^h$16$^m$24$^s$ & -16$\arcdeg$51$\arcmin$00$\arcsec$ & 13.998 & -0.128 & 40.3 & 3.65 \\
N22 & 18$^h$25$^m$10$^s$ & -13$\arcdeg$10$\arcmin$00$\arcsec$ & 18.26 & -0.3 & 51.3 & 4.0 \\
N74 & 19$^h$03$^m$55$^s$ & +5$\arcdeg$07$\arcmin$00$\arcsec$ & 38.909 & -0.437 & 40.4 & 2.8\tablenotemark{d} \\
\enddata
\tablenotetext{a}{from Churchwell et al. (2006)}
\tablenotetext{b}{from Beaumont \& Williams (2010)}
\tablenotetext{c}{from Deharveng et al. (2010)}
\tablenotetext{d}{Could be at far kinematic distance of 10.4 kpc; see
  section \ref{sec-N74}.}
\end{deluxetable}

\subsection{CARMA Observations}

CARMA's correlator allows simultaneous observation in 8 independently customizable frequency windows, each of which records data from both lower and upper side bands spaced between 1 and 9 GHz from a specified local oscillator frequency. This allowed for great flexibility in observing the thermal dust continuum and multiple spectral lines at bandwidths and spectral resolutions that would efficiently probe many components of the molecular cloud surrounding the bubbles and reveal the velocity structure of those components. All of the CARMA observations included in this paper were taken with the antennas in the D configuration, with baselines ranging from 11 to 148 meters. At 91 GHz, this corresponds to a resolution of $\sim$5$\arcsec$, and has sensitivity to scales up to about an arcminute. At the distance of the bubbles, this probes size scales of  $\sim$0.1 to 1 pc, which covers both potential star-forming ``cores'' and larger fragmented ``clumps'' and resolves out emission extended over scales that are significant fractions of the size of the bubbles, which are each a few parsecs in span.

We observed seven spectral lines in N14 as well as 3 mm continuum. $^{13}$CO (J=1$\rightarrow$0) and C$^{18}$O (J=1$\rightarrow$0) trace the overall density of the molecular gas and are generally optically thin enough to derive column densities from. CS (J=2$\rightarrow$1), being a less abundant molecule with a higher critical density, is generally seen in only denser pockets of gas. CH$_3$OH (J=8$\rightarrow$7) masers are Class I masers only excited by outflows in early stages of star formation \citep{Cyg09}, so finding masers would be a good confirmation of current star-formation activity. SiO (J=2$\rightarrow$1) is excited by shocks, so small-scale SiO emission would indicate whether star-forming cores had evolved to the point of driving their own supersonic outflows. \nh ~(J=1$\rightarrow$0) will be found in cold dark clouds, as the molecule is destroyed in high-radiation environments \citep{Womack92}, while \hco ~(J=1$\rightarrow$0) traces the dynamics of the gas and is often bright in regions of active star formation \citep{Cyg11}.

For N22 and N74, we observed a smaller selection of lines that we felt would tell us the most about the star-formation activity. CS, \nh, and \hco ~are the same as in N14, and we added HCN (J=1$\rightarrow$0) as well, which is a tracer of dense regions in the molecular cloud likely to have active star-formation \citep{Miura10}. The HCN line is also at lower frequency than the CO isotopes, so can be observed simultaneously with the other lines in a single correlator set-up.

The observations are summarized in Table \ref{table-obs1}, including the calibrators used and spatial resolution achieved. The observations of N14 used two different correlator configurations in order to observe lines too widely spread in frequency to be observed simultaneously, and also devote sufficient bandwidth to the continuum. The observed molecular lines, channel widths, and noise levels are listed in Table \ref{table-obs2}.

N14 was observed over multiple tracks in August 2010. Some observing time at the lower frequency setting was lost due to bad weather and bad passband calibration. The observations of N22 and N74 were both taken in June 2011. The data reduction and calibration was done with the MIRIAD data reduction package \citep{MIRIAD}. The noise levels and beam shapes in Table \ref{table-obs1} are derived from using ``natural weighting'' on the visibility data, which enhances small-scale structures and increases signal-to-noise at the expense of angular resolution and sidelobe suppression.

\begin{deluxetable}{cccccccccc}
\rotate
\footnotesize
\tablecaption{Summary of Observations
\label{table-obs1}
}
\tablewidth{0pt}
\tabletypesize{\scriptsize}
\tablehead{
Source & Central Freq. & Cont. bandwidth  & Lines observed &
\# pointings & Int. time & Passband Calibrator & Flux Calibrator & Phase
Calibrator & Beam size
\\
   & (GHz) & (GHz) & & & (hours) & & & & (arcseconds)
}
\startdata
N14 (low) & 91.2 & 6.0 & SiO, HCO$^+$, N$_2$H$^+$ & 50 & 7.54 & 3C454.3 & Neptune,
Mars & 1733-130 & 7.715 $\times$ 5.036 \\
N14 (high) & 102.5 & 5.0 & CS, CH$_3$OH, C$^{18}$O, $^{13}$CO & 50 & 13.3 & 3C454.3
& Neptune, Mars & 1733-130 & 6.747 $\times$ 4.653 \\
N22 & 91.2 & 5.0 & HCN, HCO$^+$, N$_2$H$^+$, CS & 66 & 11.5 & 3C454.3, 3C273 & Neptune & 1733-130 & 6.513
$\times$ 5.099 \\
N74 & 91.2 & 5.0 & HCN, HCO$^+$, N$_2$H$^+$, CS & 58 & 12.07 & 3C454.3 & Neptune & 1751+096 & 5.231 $\times$ 4.961 \\
\enddata

\end{deluxetable}
\begin{deluxetable}{cccccc}
\footnotesize
\tablecaption{Molecular Lines Observed
\label{table-obs2}
}
\tablewidth{0pt}
\tabletypesize{\scriptsize}
\tablehead{
Source & Molecule & Transition  & Frequency & Channel width & Noise level 
\\
   &  &   & (GHz) & (km s$^{-1}$) & (mJy beam$^{-1}$) 
}
\startdata
N14 & SiO & J=2$\rightarrow$1 & 86.24352 & 0.566 & 0.117 \\
N14 & HCO$^+$ & J=1$\rightarrow$0 & 89.18852 & 0.547 & 0.147 \\
N14 & N$_2$H$^+$ & J=1$\rightarrow$0 & 93.17351 & 0.524 & 0.149 \\
N14 & CS & J=2$\rightarrow$1 & 97.98097 & 0.249 & 0.142 \\
N14 & CH$_3$OH & J=8$\rightarrow$7 & 95.16944 & 0.513 & 0.115 \\
N14 & C$^{18}$O & J=1$\rightarrow$0 & 109.78216 & 0.444 & 0.119 \\
N14 & $^{13}$CO & J=1$\rightarrow$0 & 110.20135 & 0.443 & 0.133 \\
\hline
N22 & HCN & J=1$\rightarrow$0 & 88.63185 & 0.330 & 0.096 \\
N22 & HCO$^+$ & J=1$\rightarrow$0 & 89.18852 & 0.328 & 0.111 \\
N22 & N$_2$H$^+$ & J=1$\rightarrow$0 & 93.17351 & 0.314 & 0.109 \\
N22 & CS & J=2$\rightarrow$1 & 97.98097 & 0.299 & 0.111 \\
\hline
N74 & HCN & J=1$\rightarrow$0 & 88.63185 & 0.330 & 0.044 \\
N74 & HCO$^+$ & J=1$\rightarrow$0 & 89.18852 & 0.328 & 0.053 \\
N74 & N$_2$H$^+$ & J=1$\rightarrow$0 & 93.17351 & 0.314 & 0.056 \\
N74 & CS & J=2$\rightarrow$1 & 97.98097 & 0.299 & 0.051 \\
\enddata

\end{deluxetable}

\subsection{Survey Data}

We compared our data with public survey data from four galactic-plane surveys at wavelengths ranging from 3 $\micron$ to 20 cm, using the NASA/IPAC Infrared Science Archive (IRSA) to extract ``cutouts'' of the regions around the three bubbles for each of the surveys.

The original Ch06 catalog from which these sources are drawn identified the bubbles from the Spitzer-GLIMPSE survey. This survey covers the inner 130 degrees of the galactic plane and galactic latitudes from -1$^{\circ}$ to 1$^{\circ}$. The survey used the IRAC instrument \citep{IRAC} in four wavelength bands ranging from 3.6 to 8.0 $\micron$. The band of greatest interest in this paper is the 8.0-$\micron$ band, as it contains polycyclic aromatic hydrocarbon (PAH) emission at 7.7 and 8.6 $\micron$ that are luminous in the PDR surrounding \hii ~regions. In this band, Spitzer has a resolution of $\sim$2$\arcsec$.

The Spitzer-MIPSGAL survey covers the same area with the MIPS instrument \citep{MIPS} at 24 \micron ~with 6$\arcsec$ resolution. MIPSGAL revealed diffuse 24-\micron ~emission, dominated by continuum emisision from warm dust, inside many of the GLIMPSE bubbles that seems to trace the region inside the PDR. Around the three bubbles discussed in this paper, MIPSGAL also revealed a number of point sources, which, as they coincide with 8 \micron ~point sources with varying ratios of flux at the two wavelengths, we interpret as protostars of different ages.

The Multi-Array Galactic Plane Imaging Survey (MAGPIS; \citealt{MAGPIS}) is a 20 cm continuum survey of the northern galactic plane with the VLA. The large scale structure that the VLA resolved out was included by adding data from the 100-m Effelsberg telescope. The survey covers galactic longitudes of 5$^{\circ}$ to 48$^{\circ}$ and latitudes from -0.8$^{\circ}$ to 0.8$^{\circ}$ with a resolution of 5$\arcsec$. This data is largely used to detect free-free emission from \hii ~regions. The \hii ~regions traced by this survey are another part of the large-scale picture of the environment of the bubbles. The coincidence of drop-offs in 20 cm data with high surface brightness 8 $\micron$ data were taken to draw the outline of the expanding bubble.

The BGPS survey covers a slightly irregular area of 150 square degrees which includes the northern galactic plane from 10$^{\circ}$ to 90.5$^{\circ}$ longitude and from -0.5$^{\circ}$ to 0.5$^{\circ}$ latitude, which includes a large number of GLIMPSE bubbles, including the ones discussed in this paper. The BGPS was taken with the Bolocam instrument at the Caltech Submillimeter Observatory at a wavelength of 1.1 mm, chosen to exclude CO (2$\rightarrow$1) and hence be dominated by thermal continuum emission. The instrument has a FWHM beam of 33$\arcsec$. High surface brightness in the BGPS was one of the elements in source selection, and the comparison of the dust emission seen at 3 mm by CARMA and at 1.1 mm by the BGPS will be discussed in section \ref{cha:fragmentation}.

\begin{figure}
\begin{center}
\includegraphics[height=2.25in]{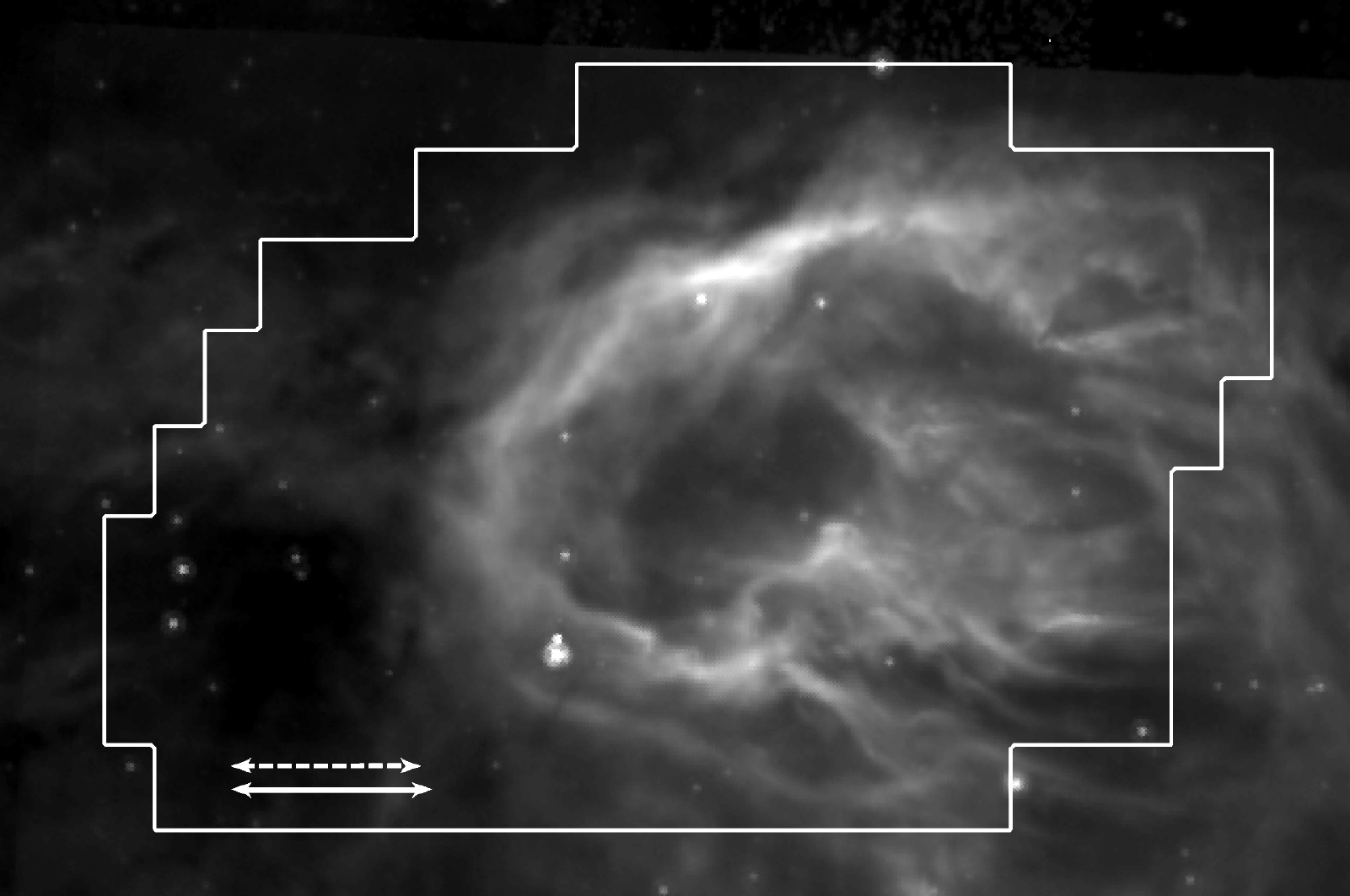}
\includegraphics[height=2.25in]{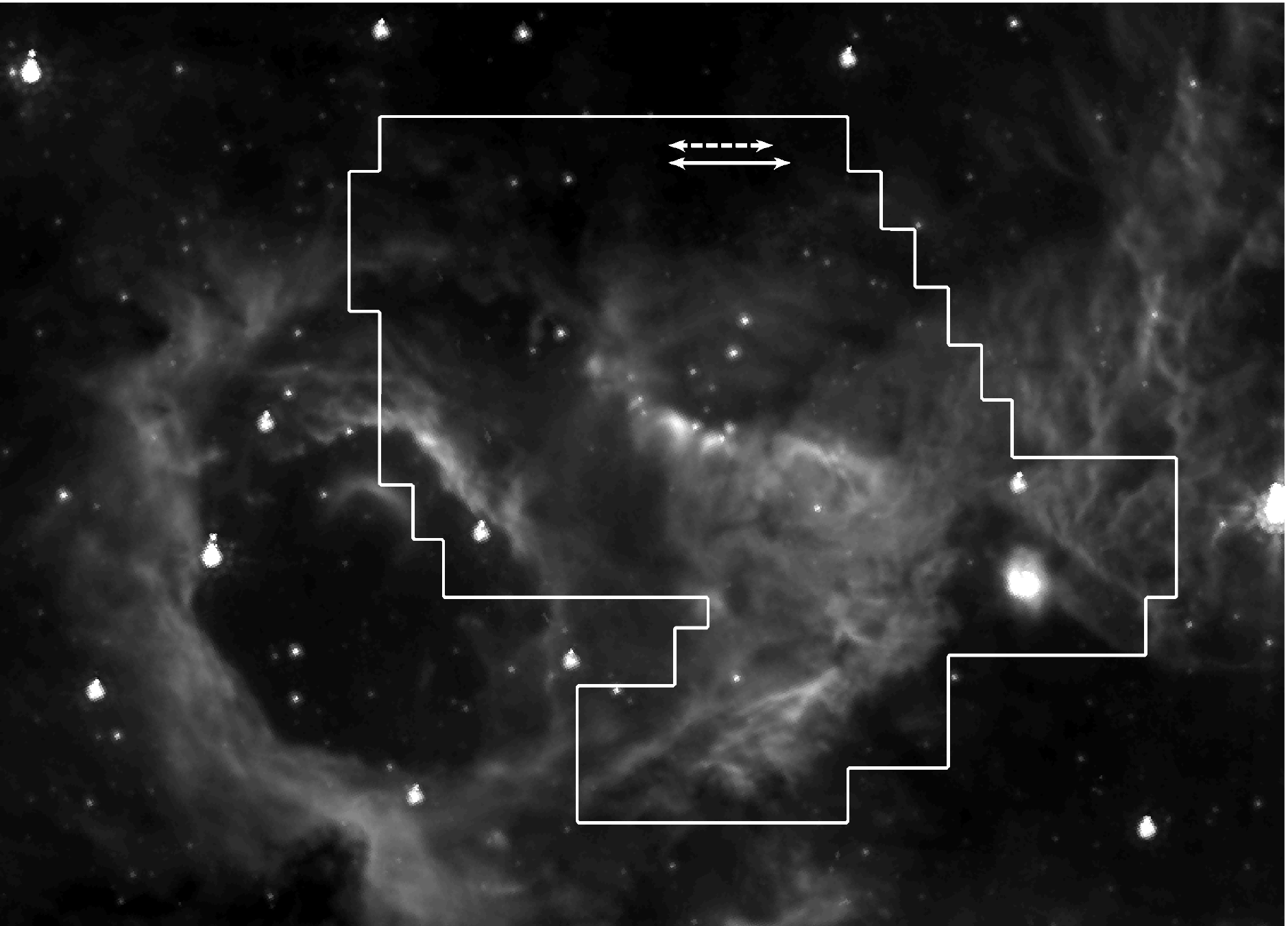}
\includegraphics[height=2.25in]{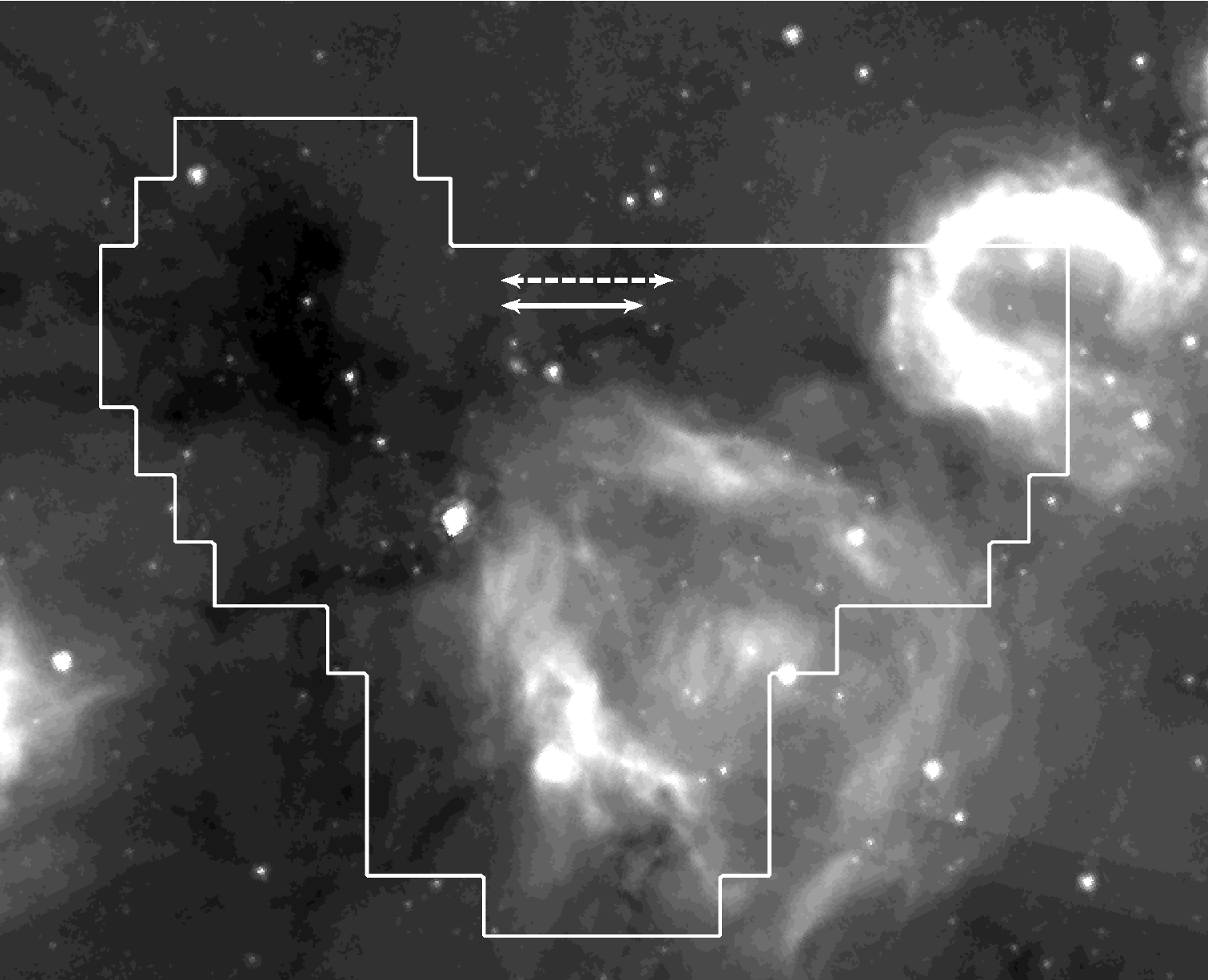}
\end{center}
\caption[Bubble Fields of View] {N14 (top), N22 (bottom left), and
  N74 (bottom right). 8 $\micron$ GLIMPSE data is in grayscale. The white
  polygons show the areas mapped by CARMA. The solid arrow is an
  arcminute in length for scale reference, and the dashed arrow is a
  parsec in length at the assumed distance of each bubble.
\label{fig-bubblecolor}}
\end{figure}

Figure \ref{fig-bubblecolor} shows the three bubbles we observed. The grayscale is 8 $\micron$ GLIMPSE data, primarily tracing PAHs in the photodissociation region, and an outline of the CARMA field of view is shown. Arrows represent reference angular and physical scales, with solid arrows of length one arcminute, and dashed arrows one parsec.

\section{RESULTS}
\label{cha:results}

Emission was detected by CARMA at locations in the vicinity of each bubble in continuum and nearly every line observed. Around N14, No SiO emission was detected and no CH$_3$OH masers were detected. The implications of this will be discussed in section \ref{cha:fragmentation}.

This paper will focus on a subset of the data. The spatial distribution of \nh ~gas will be discussed, and the properties of the small-scale structures in continuum emission will be presented. The other molecular line emission will be used to derive velocities of dust cores and to give a picture of the overall velocity structure of the bubbles and surrounding clouds, but further analysis of the fragmentation seen in the molecular line data will be presented in a future paper.

\subsection{\nh ~Clumps}

\begin{figure}
\begin{center}
\includegraphics[height=2.5in]{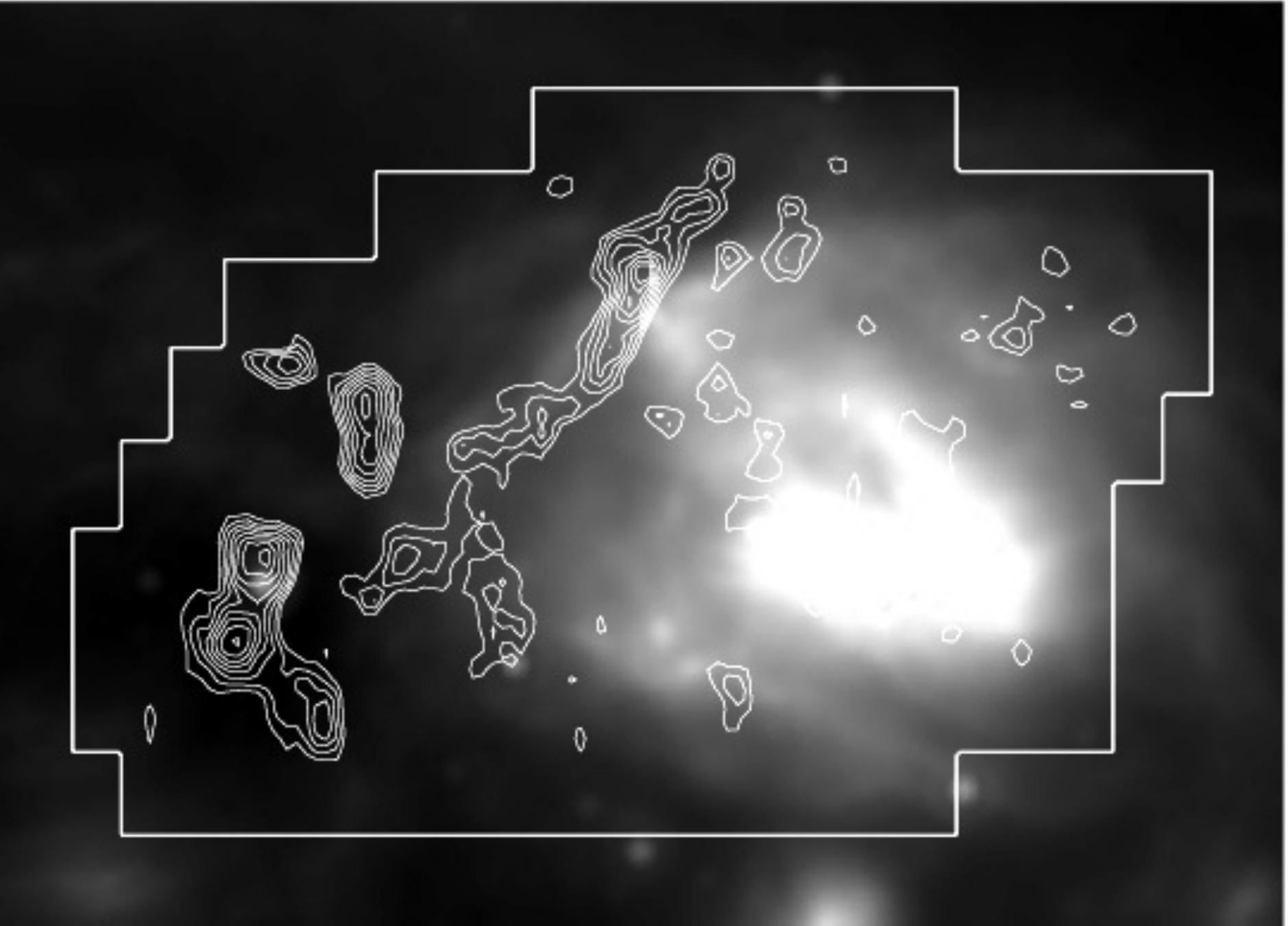}
\end{center}
\caption[N14 \nh ~Emission] {N14: 24 $\micron$ with contours of
  \nh. The \nh ~data has been integrated over velocities
  from 39.38 to 42.52 km s$^{-1}$, and smoothed with a 6$\arcsec$
  kernel.
\label{fig-N14N2H}}
\end{figure}

\begin{figure}
\begin{center}
\includegraphics[height=2.5in]{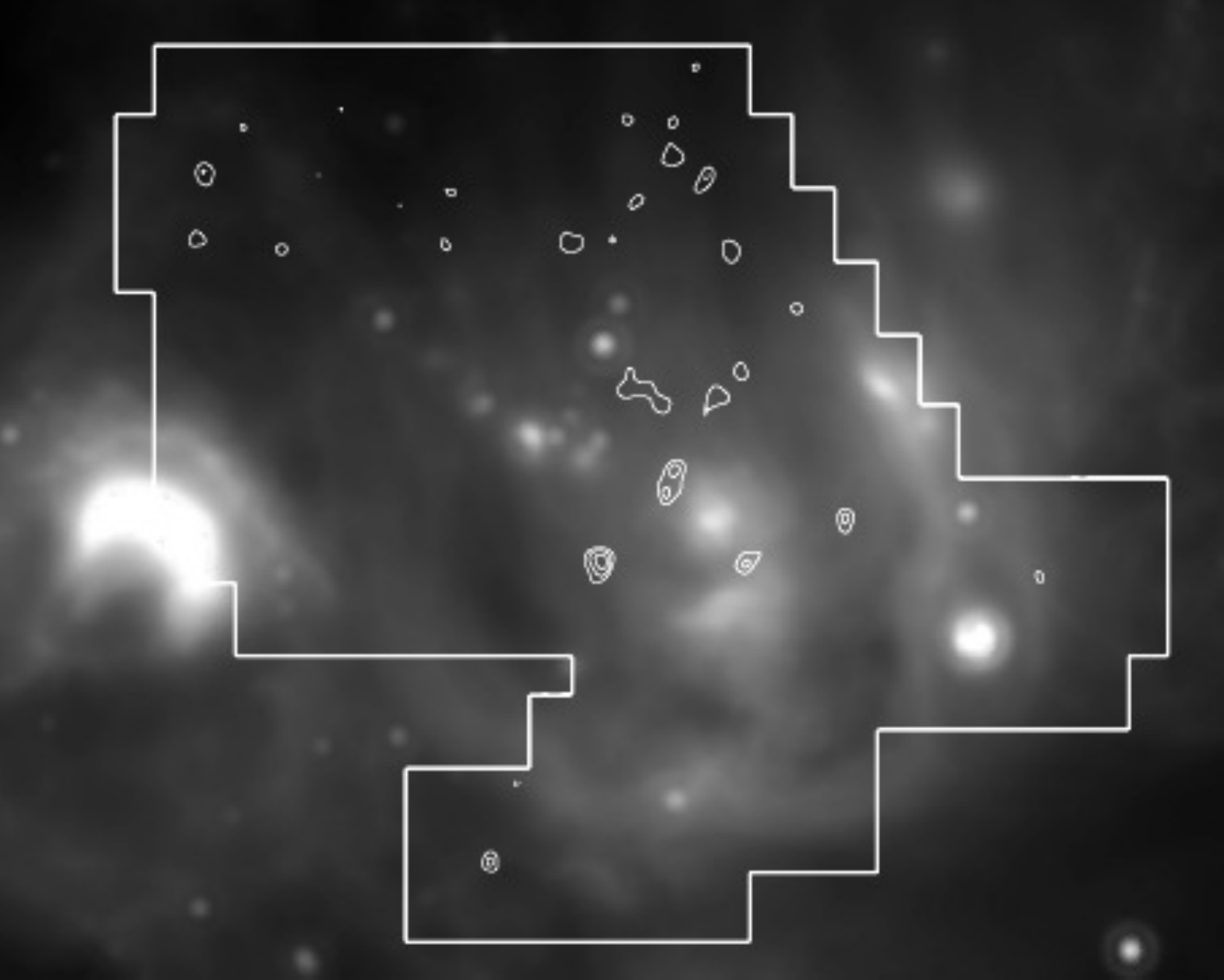}
\includegraphics[height=2.5in]{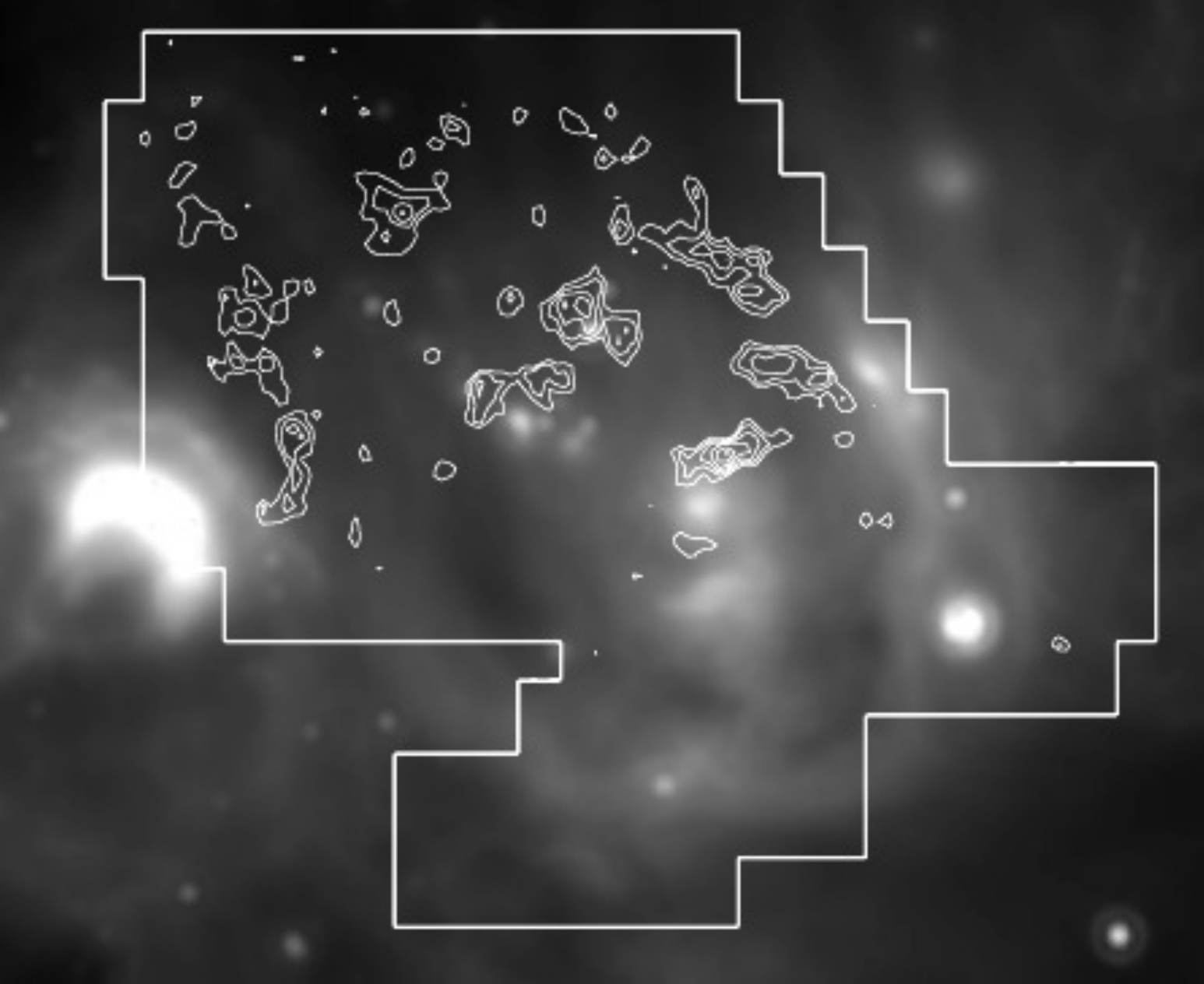}
\end{center}
\caption[N22 \nh ~Emission] {N22: 24 $\micron$ with contours of \nh. The \nh ~data has been integrated over velocities from 49.05 to 53.14 km s$^{-1}$ (left) and from 62.88 to 66.96 km s$^{-1}$ (right), and smoothed with a 6$\arcsec$ diameter kernel.
\label{fig-N22N2H}}
\end{figure}

\begin{figure}
\begin{center}
\includegraphics[height=2.5in]{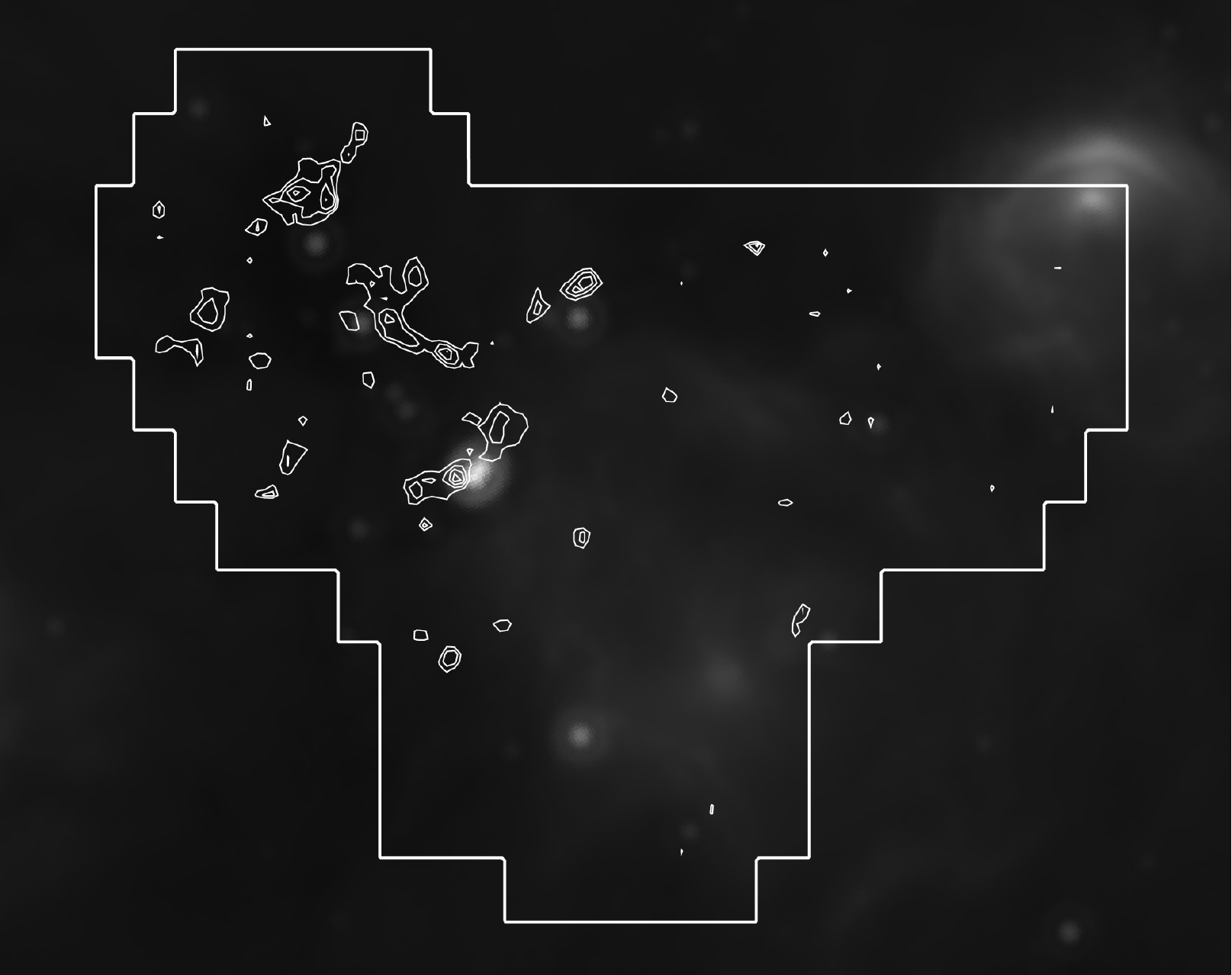}
\end{center}
\caption[N74 \nh ~Emission] {N74: 24 $\micron$ with contours of
  \nh. The \nh ~data has been integrated over velocities
  from 40.49 to 43.63 km s$^{-1}$, and smoothed with a 6$\arcsec$
  kernel.
\label{fig-N74N2H}}
\end{figure}

Figures \ref{fig-N14N2H}, \ref{fig-N22N2H}, and \ref{fig-N74N2H} show 24-$\micron$ MIPSGAL emission in grayscale with contours of \nh ~overlaid. As can be seen, clumps of \nh ~are present throughout the cloud surrounding the bubbles. Almost all the clumps, particularly the brightest ones, are seen along lines of sight that do not have extended 24 $\micron$ dust emission, which originates within the ionizing shock front. This is not surprising, as \nh ~is generally observed to be present in quiescent cold clouds but is absent in shocked gas (e.g.~\citealt{Womack92}). The fact that the \nh ~emission is well aligned with the bubble rim confirms that the velocity components we are mapping are associated with the bubble and are not coincident foreground or background clouds.

Around N14 and N74, all the \nh ~we detected was in a narrow range of velocities ($<$ 3 km s$^{-1}$). In both, there are a few clumps seen within the borders of the bubble, though fewer in number and lower in peak flux than those outside the bubble.

Around N22, \nh ~is seen in two distinct velocity components, centered at 51 km s$^{-1}$ and 65 km s$^{-1}$. The \hii region that N22 encloses, which is at the east edge of the CARMA field of view, is at a velocity of 50.9 km s$^{-1}$ \citep{Anderson09b}, so the emission at 51 km s$^{-1}$ is almost certainly associated with the bubble region. The 65 km s$^{-1}$ emission is confined to the northern half of the field mapped by CARMA, and could plausibly originate in the gas seen as extinction against the bright infrared background. Whether this is the case, and whether this cloud is interacting in some way with the bubble system or whether it is just in the line of sight is ambiguous.

\subsection{Dust Continuum}

In Table 2 of D10, they list dust ``condensations'' they have found around bubbles using the APEX Telescopse Large Area Survey of the Galactic plane at 870 $\micron$ (ATLASGAL; \citealt{ATLASGAL}), which has resolution of 19$\farcs$2. They discovered three around N14, four around N22, and none around N74. They do not name them; we will refer to them as D-1, D-2, etc., numbered separately around each bubble.

To identify dust sources, we used the 102 GHz data from N14, as we had longer integration time, and the dust emission is brighter at the higher frequency. From the 3-mm continuum data, we have identified 28 compact dust sources: ten in the vicinity of N14, fourteen around N22, and four around N74. To distinguish between true sources and noise peaks, we smoothed the data with a 9$\arcsec$-diameter Gaussian beam and set a cut-off of 3-$\sigma$. The smoothing cuts down on places where the noise reaches the 3-$\sigma$ level over a pixel or two, but does not overly diffuse sources that are unresolved by our $\sim$5$\arcsec$ beam. We disregarded peaks closer than 20$\arcsec$ to the edge of the observed field, as points along that border were not observed by as many overlapping pointings, and hence have higher noise. All the sources we identify are co-located with molecular line emission and are consistent in spatial distribution with the 1.1-mm BGPS data, which provides confirmation that they are real.

We will refer to these sources as ``cores'' due to their compact size. As will be discussed in section \ref{cha:bubbles}, some could be small filaments aligned along the line of sight with a total column density high enough to be detected as an unresolved source by CARMA rather than spherical objects. They could also be the peaks of extended clumps, with much of the extended emission resolved out by our interferometric data. For the rest of the paper, we will use ``core'' for shorthand to refer to our compact sources.

Table \ref{table-dustcores} summarizes the properties of the dust cores, including their fluxes, estimated masses, and velocities. The formula used to estimate the masses is discussed in section \ref{sec:mass}, and the associated molecular line emission in section \ref{sec:velocity}. Velocities were derived from the \hco ~line, though in a few cases distinguishing between components was done with other lines.

\begin{deluxetable}{cccccccccc}
\footnotesize
\tablecaption{3.3-mm Continuum Dust Cores
\label{table-dustcores}
}
\tablewidth{0pt}
\tabletypesize{\scriptsize}
\tablehead{
Bubble & Core &$\alpha$(J2000)  & $\delta$(J2000) & Morphology\tablenotemark{a} & Flux &
Mass & Velocity & 24 $\micron$ Point Source? & Associations \\
  &   &  &  & (mJy) & (M$_\sun$) & (km s$^{-1}$) & &}

\startdata
N14 & A & 18$^h$16$^m$17$\fs$32 & -16$\arcdeg$50$\arcmin$13$\farcs$0 &
C & 3.2 & 36 & 39.7 & No & Northern rim \\
N14 & B & 18$^h$16$^m$18$\fs$01 & -16$\arcdeg$50$\arcmin$54$\farcs$3 &
C & 4.0 & 45 & 42.4\tablenotemark{b} & No & Bubble interior \\
N14 & C & 18$^h$16$^m$18$\fs$41 & -16$\arcdeg$50$\arcmin$46$\farcs$7 &
M & 7.1 & 80 & 42.4\tablenotemark{b} & No & Bubble interior \\
N14 & D & 18$^h$16$^m$21$\fs$51 & -16$\arcdeg$51$\arcmin$24$\farcs$4 &
M & 7.0 & 79 & 42.6 & No & 8-$\micron$ pillar \\
N14 & E & 18$^h$16$^m$22$\fs$46 & -16$\arcdeg$51$\arcmin$58$\farcs$7 &
M & 24.3 & 274 & 38.8 & No & Southern rim, D-1 \\
N14 & F & 18$^h$16$^m$23$\fs$35 & -16$\arcdeg$50$\arcmin$25$\farcs$8 &
M & 21.9 & 247 & 41.2\tablenotemark{c} & No & Bubble interior \\
N14 & G & 18$^h$16$^m$24$\fs$77 & -16$\arcdeg$49$\arcmin$52$\farcs$3 &
E & 21.2 & 239 & 42.3 & No & Northern rim, D-2 \\
N14 & H & 18$^h$16$^m$25$\fs$86 & -16$\arcdeg$51$\arcmin$32$\farcs$4 &
C & 3.4 & 38 & ? & No & Bubble interior \\
N14 & I & 18$^h$16$^m$30$\fs$25 & -16$\arcdeg$50$\arcmin$37$\farcs$8 &
M & 14.6 & 165 & 42.6 & No & Northeast rim \\
N14 & J & 18$^h$16$^m$33$\fs$08 & -16$\arcdeg$51$\arcmin$23$\farcs$3 &
E & 21.8 & 246 & 42.3 & Yes & IRDC, D-3 \\
\hline
N22 & A & 18$^h$24$^m$53$\fs$92 & -13$\arcdeg$09$\arcmin$52$\farcs$4 &
M & 10.0 & 213 & 51.2 & No & PDR \\
N22 & B & 18$^h$24$^m$55$\fs$10 & -13$\arcdeg$10$\arcmin$00$\farcs$5 &
C & 4.8 & 102 & 55.2 & No & IRDC \\
N22 & C & 18$^h$24$^m$55$\fs$25 & -13$\arcdeg$09$\arcmin$45$\farcs$6 &
C & 2.6 & 55 & 52.9 & No & PDR \\
N22 & D & 18$^h$24$^m$55$\fs$84 & -13$\arcdeg$10$\arcmin$18$\farcs$0 &
C & 8.4 & 179 & 55.2\tablenotemark{d} & Yes & IRDC,YSO \\
N22 & E & 18$^h$25$^m$01$\fs$47 & -13$\arcdeg$09$\arcmin$47$\farcs$5 &
M & 11.7 & 249 & 51.9 & No & PDR \\
N22 & F & 18$^h$25$^m$02$\fs$08 & -13$\arcdeg$09$\arcmin$15$\farcs$8 &
E & 14.5 & 309 & 51.8\tablenotemark{e} & No & PDR, D-1 \\
N22 & G & 18$^h$25$^m$03$\fs$68 & -13$\arcdeg$09$\arcmin$52$\farcs$3 &
M & 7.3 & 156 & 51.2 & No & PDR \\
N22 & H & 18$^h$25$^m$03$\fs$82 & -13$\arcdeg$09$\arcmin$31$\farcs$8 &
C & 3.1 & 66 & 51.3 & No & PDR \\
N22 & I & 18$^h$25$^m$04$\fs$49 & -13$\arcdeg$08$\arcmin$31$\farcs$0 &
E & 17.7 & 377 & 65.7 & No & YSO cluster \\
N22 & J & 18$^h$25$^m$05$\fs$66 & -13$\arcdeg$08$\arcmin$23$\farcs$9 &
C & 17.9 & 381 & 65.8 & Yes & YSO cluster,D-2 \\
N22 & K & 18$^h$25$^m$06$\fs$00 & -13$\arcdeg$08$\arcmin$06$\farcs$7 &
C & 2.6 & 55 & 66.1 & Yes & YSO cluster \\
N22 & L & 18$^h$25$^m$06$\fs$53 & -13$\arcdeg$08$\arcmin$54$\farcs$2 &
M & 14.5 & 309 & 65.8\tablenotemark{f} & Yes & YSO cluster,D-3 \\
N22 & M & 18$^h$25$^m$11$\fs$86 & -13$\arcdeg$08$\arcmin$05$\farcs$6 &
C & 34.4 & 733 & 65.6 & No & YSO cluster,D-4 \\
N22 & N & 18$^h$25$^m$14$\fs$73 & -13$\arcdeg$06$\arcmin$52$\farcs$7 &
C & 3.1 & 66 & 66.3\tablenotemark{c} & No & Northern cloud \\
\hline
N74 & A & 19$^h$04$^m$02$\fs$81 & +05$\arcdeg$07$\arcmin$58$\farcs$8 &
C & 4.1 & 43/590\tablenotemark{g} & 40.6 & No & Bubble rim \\
N74 & B & 19$^h$04$^m$03$\fs$47 & +05$\arcdeg$07$\arcmin$54$\farcs$6 &
M & 9.0 & 94/1297 & 40.9 & Yes & Bubble rim \\
N74 & C & 19$^h$04$^m$07$\fs$25 & +05$\arcdeg$09$\arcmin$44$\farcs$6 &
C & 6.4 & 67/922 & 44.4 & No & IRDC \\
N74 & D & 19$^h$04$^m$07$\fs$54 & +05$\arcdeg$08$\arcmin$46$\farcs$0 &
C & 5.1 & 53/735 & 40.6 & No & IRDC \\
\enddata

\tablenotetext{a}{C: compact; less than twice the beamshape in
  size. M: multiple peaks not distinguished by CARMA's resolution. E:
  extended flux surrounding a single peak.}
\tablenotetext{b}{Continuum peak is $\sim$8$\arcsec$ northwest of \hco ~emission.}
\tablenotetext{c}{Continuum peak is $\sim$5$\arcsec$ east of \hco ~emission.}
\tablenotetext{d}{Continuum peak is $\sim$10$\arcsec$ southeast of
  \hco ~emission.}
\tablenotetext{e}{No \hco ~peak present, so velocity measured from HCN
  emission.}
\tablenotetext{f}{\hco ~also seen at 51 km s$^{-1}$, but 65.8
  km s$^{-1}$ gas is more compact and luminous in HCN, so we
  interpret that as being associated with dust core.}
\tablenotetext{g}{\textrm{L}ow masses assume near kinematic distance of 2.8 kpc, high masses assume far kinematic distance of 10.4 kpc.}
\end{deluxetable}

In N14, six of the ten dust cores seem to be located in the PDR, as they are co-located with intense 8-$\micron$ emission. Two are in the interior of the bubble, and two are part of a dense ridge of molecular gas just to the east of N14. Our source E is likely associated with D-1, G is co-located with D-2, and J is co-located with D-3.

N22's morphology is complicated, and will be discussed more in section \ref{sec-N22}, but at least six of the fourteen cores seem to be associated with an extended cloud of 8-$\micron$ emission, while six are more likely associated with a molecular cloud to the north, especially when the velocity information is included. At least three of the four dust condensations from D10 all likely originate in the molecular cloud. D-1 is very close to our 3.3-mm core F, while D-2 is our source J, D-3 is source L, and D-4 is just west of our source M.

In N74, two of the four cores are located at the rim of the bubble, while the other two originate in the IRDC extended to the northeast of the bubble.

\begin{figure}
\begin{center}
\includegraphics[width=5.75in]{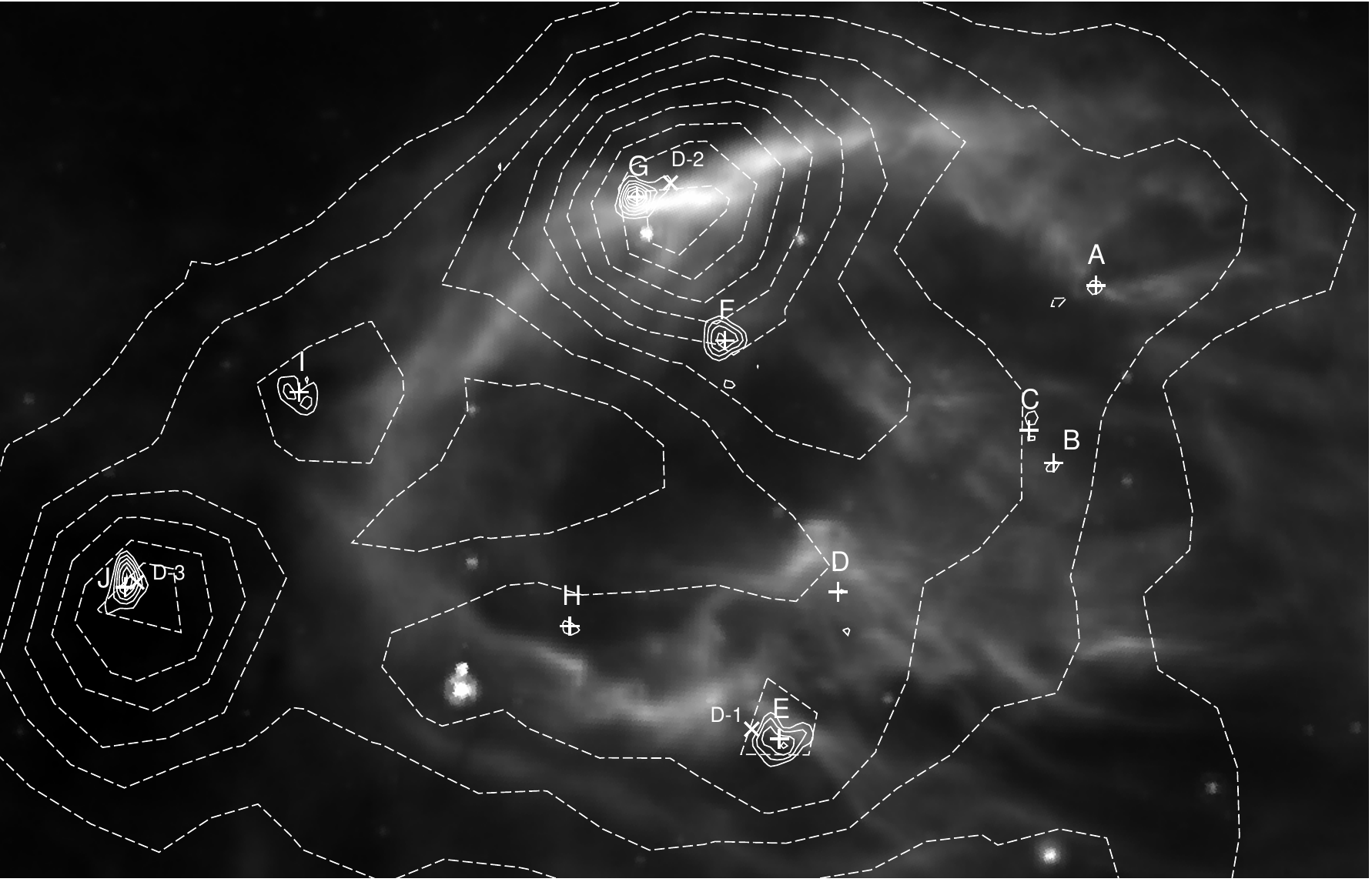}
\end{center}
\caption[N14 Dust Cores] {N14 GLIMPSE 8 $\micron$ image in
  grayscale. Labeled crosses are the 2.9 mm dust cores found by CARMA. Dust
  condensations listed in D10 are labeled with X's. Contours are of
  BGPS 1.1 mm data.
\label{fig-N14dust}}
\end{figure}

\begin{figure}
\begin{center}
\includegraphics[width=5.75in]{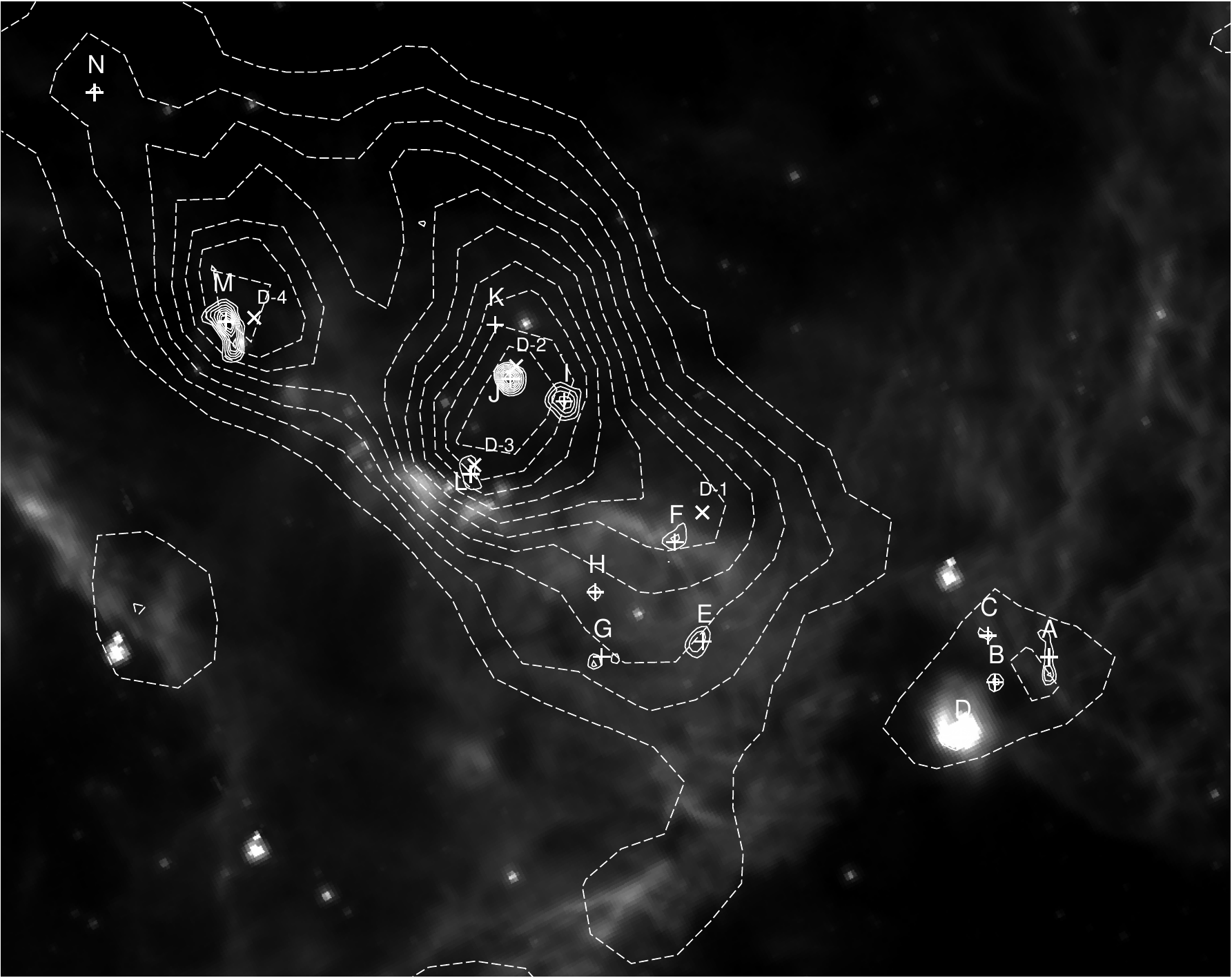}
\end{center}
\caption[N22 Dust Cores] {N22 GLIMPSE 8 $\micron$ image in
  grayscale. Labeled crosses are the 3.3 mm dust cores found by CARMA. Dust
  condensations listed in D10 are labeled with X's. Contours are of
  BGPS 1.1 mm data.
\label{fig-N22dust}}
\end{figure}

\begin{figure}
\begin{center}
\includegraphics[width=5.75in]{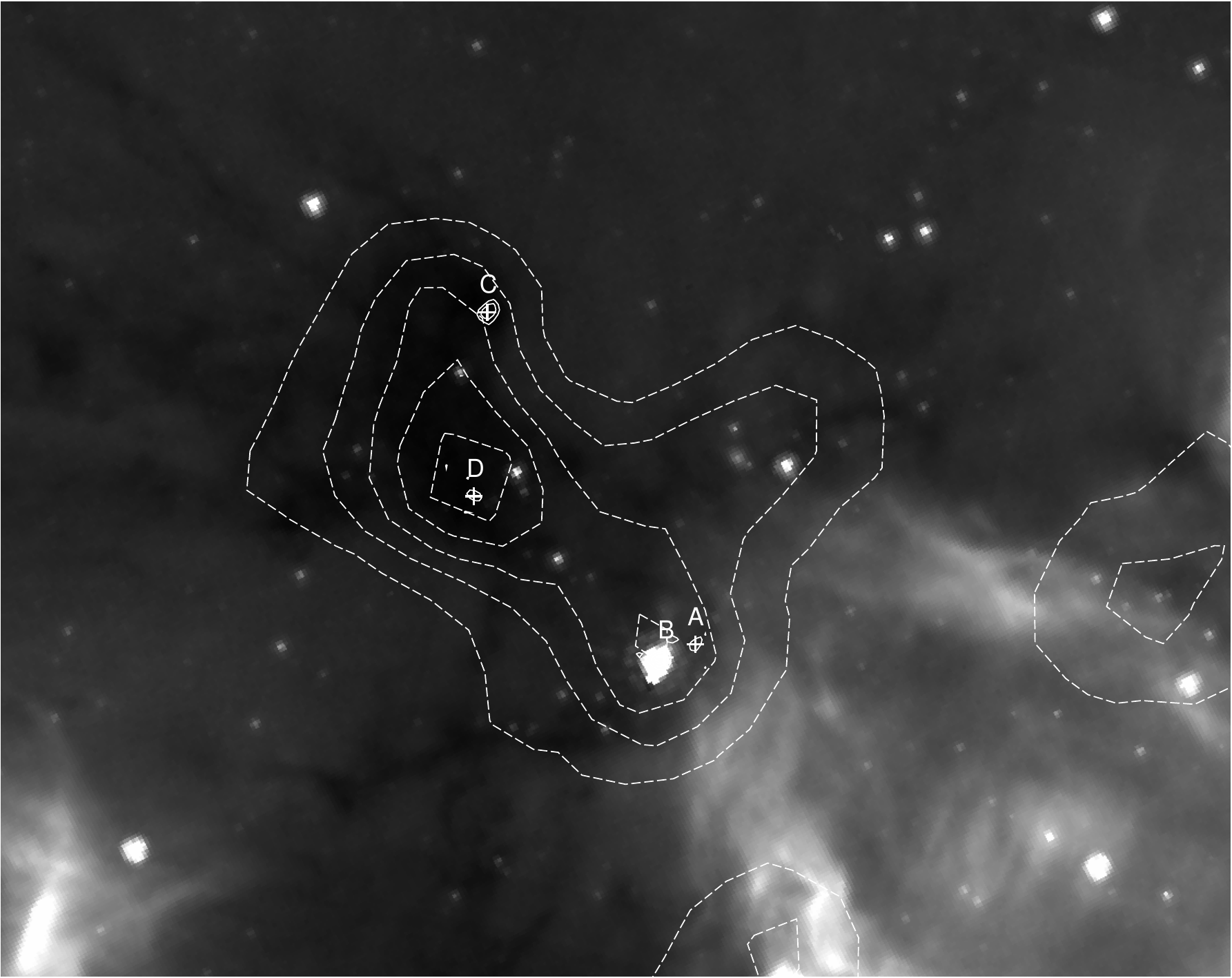}
\end{center}
\caption[N74 Dust Cores] {N74 GLIMPSE 8 $\micron$ image in
  grayscale. Labeled crosses are the 3.3 mm dust cores found by CARMA. Contours are of
  BGPS 1.1 mm data.
\label{fig-N74dust}}
\end{figure}

Figures \ref{fig-N14dust}, \ref{fig-N22dust}, and \ref{fig-N74dust} show the 8-micron GLIMPSE images of N14, N22, and N74 with symbols showing the placement of the 3-mm dust cores. The dust cores are crosses labeled with their designation, and the locations of the dust condensations listed in D10 are labeled with X's. BGPS data is shown in contours.

\subsubsection{Dust core mass estimates}
\label{sec:mass}

For each dust core we identified, we fit a two-dimensional Gaussian to the area around the peak to determine some of the source properties. Because the signal-to-noise of many individual dust cores is fairly small, our uncertainties on source size and total flux were large. The uncertainties on source position were smaller than the pixel size in our CARMA maps of 1$\farcs$5, so we took these positions as being the true source center.

Many of the source shapes were not well-fit by a Gaussian, and some (as noted in Table \ref{table-dustcores}) seem to be either extended or contain multiple unresolved peaks. To estimate the fluxes, we summed the emission surrounding the peaks down to the 3-$\sigma$ level. The fluxes measured by this method are listed in Table \ref{table-dustcores}. As well as statistical uncertainties, the background level may fluctuate across the image on scales too small to measure accurately, due to extended dust emission, so the fluxes are only certain to within $\sim$50\%.

Deriving masses of the dust cores requires making a number of assumptions. The first assumption is that the emission is entirely due to thermal dust and not free-free emission. Free-free emission is seen in hot ionized gas and could potentially be coming from the outer layers of dust cores irradiated by the massive stars driving the \hii region, as is seen in the Orion Nebula \citep{Odell93}. The dust cores we are mapping are generally in the shell outside the \hii region, though, so we do not expect free-free emission to be significant. We can confirm this with the MAGPIS data, which is at similar resolution to our CARMA data but at a wavelength where free-free emission should be stronger. If any of our dust cores had significant free-free components in their spectra at 3 mm, we would be able to measure that flux at 20 cm. We do not see any bright 20 cm spots at the locations of our dust cores, so while a few have low-level diffuse 20 cm emission in the area, any free-free emission at 3.3 mm will just alter the background level, which we already attempted to factor out.

The second assumption is that the dust is at a single temperature. This may not be true if there is an internal heating source or one side is exposed to the high-radiation environment inside the bubble. However, at 3 mm, we are near the Rayleigh-Jeans regime, so that our derived mass will be approximately linear with assumed temperature. If we pick a plausible average temperature, this will not affect our results by a large factor. Once we assume isothermal dust and no significant free-free contribution, the mass can be estimated by the equation

\begin{equation}
M = \frac{S_{\nu} d^2}{\kappa_{\nu} B_{\nu}(T)}.
\label{eq:dustmass}
\end{equation}

In equation (\ref{eq:dustmass}), $S_{\nu}$ is the integrated flux density, $d$ is the distance, $\kappa_{\nu}$ is the dust emissivity at 3.3 mm (or 2.9 mm for the N14 cores), and $B_{\nu}(T)$ is the Planck function at dust temperature $T$. We have to assume values for $\kappa_{\nu}$ and $T$. Dust emissivity is generally expressed as $\kappa_{\nu} = \kappa_0 (\nu/\nu_0)^{\beta}$ \citep{Hildebrand83}. We will use  $\kappa_{230~GHz}$ = 0.005 cm$^2$ g$^{-1}$. \citet{Anderson10} studied the SEDs of submillimeter sources around the \hii region RCW 120 and found approximate average values of $T$ = 20 K and $\beta$ = 2.0. As our dust cores are in a similar environment, we will assume similar values. The masses derived using this equation are listed in Column 6 of Table \ref{table-dustcores}.

With so many assumptions, these estimated masses are far from exact. If any of the cores are developing into massive protostars, they may have a temperature closer to 30 K \citep{Mol00}, though this would only change the masses by $\sim$40\%. The bigger uncertainty lies in $\kappa_{\nu}$, which, due to uncertainties in both $\kappa_0$ and $\beta$ \citep{Pollack94}, could be off of our estimation by a factor of two or more. As the cores are of comparable masses and arise in somewhat similar environments, the hope is that whatever variation exists is not drastic among the objects our sample. While the three bubbles have measured kinematic distances, the distances are uncertain to $\sim$30\%, which could throw our estimated masses off by 60\%. Even with all these unknowns, our estimated masses give an order of magnitude estimate for the mass of gas in dense fragments in the clouds surrounding these bubbles. They also provide a first guess for parameters of these potential star-forming cores that can be determined in greater detail with further observation.

\subsubsection{Velocity structure}
\label{sec:velocity}

To find the velocities of the dust cores, we investigated the spectra and morphology of our observed molecular lines at the location of each dust core.

\begin{figure}
\begin{center}
\includegraphics[width=6.0in]{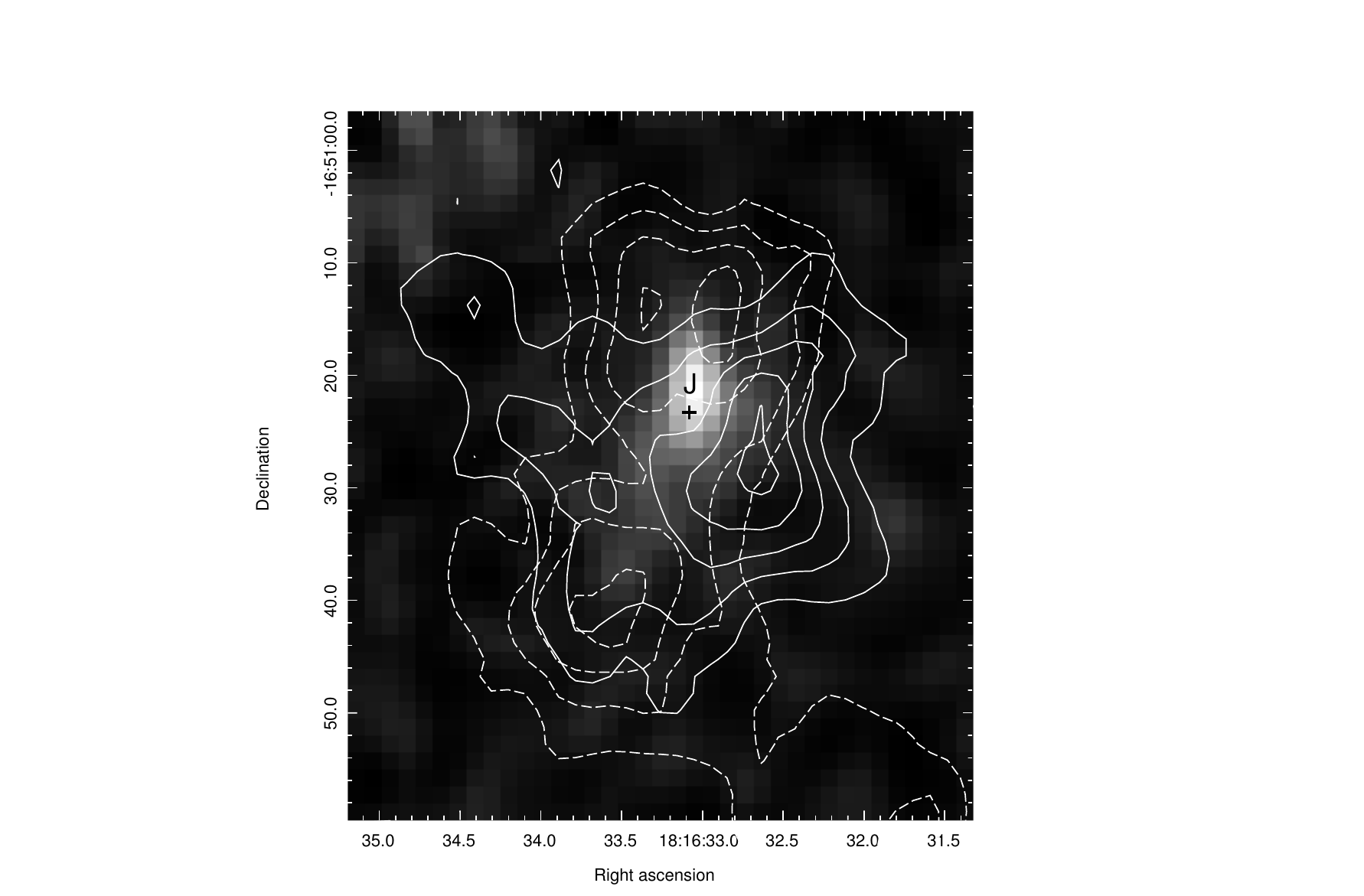}
\end{center}
\caption[\hco ~and \nh ~around N14-E] {Grayscale image of 2.9 mm
  continuum of the emission east of N14. Solid contours are
  velocity-integrated \hco ~and dashed contours are
  velocity-integrated \nh. Contours of both are at levels of 0.2 up to
  1.0 mJy beam$^{-1}$, spaced by 0.2. Dust core N14-J is labeled with
  a cross.
\label{fig-N14EFG}}
\end{figure}

\begin{itemize}

\item
Only five of the 28 sources had associated CS emission. \citet{Tafalla04} found that CS is depleted in starless cores, as it freezes onto dust grains, so this is not surprising. Three of the five are associated with a cluster of protostars seen in 24 $\micron$, so the CS emission may be from an envelope around an accreting protostar, which is sometimes detected in CS (e.g. \citealp{WC12}).

\item
In N14, $^{13}$CO and C$^{18}$O are seen in the direction of most of the dust cores, but generally as part of extended clumps, so we do not interpret this gas as being tightly bound to the dust cores but rather is tracing larger clumps in the cloud surrounding the bubble.

\item
\nh ~is seen around all of the dust cores in N14 and about half of the cores in N22 and N74. In many cases the \nh ~emission was adjacent to rather than coincident with the continuum peaks. \citet{Friesen10} found \nh ~depletion in dust cores, particularly star-forming cores, and our results corroborate this, most obviously in the cluster of source J in N14 (see Figure \ref{fig-N14EFG}). Because of that, and the difficulty in deconvolving the hyperfine structure lines to get velocities to greater accuracy than $\sim$1 km s$^{-1}$, we do not use \nh ~to measure the velocities of the cores.

\item
\hco ~and HCN (in the bubbles for which it was observed) emission is seen at nearly every source, often in compact knots. In some cases there are a bright clump of \hco ~offset of the continuum peak by 5-10$\arcsec$, which we still took to be associated with the core. We interpreted this tracer as most tightly coupled to the dust cores and hence derived velocities from the \hco ~line.

\end{itemize}

\begin{figure}
\begin{center}
\includegraphics[width=2.75in]{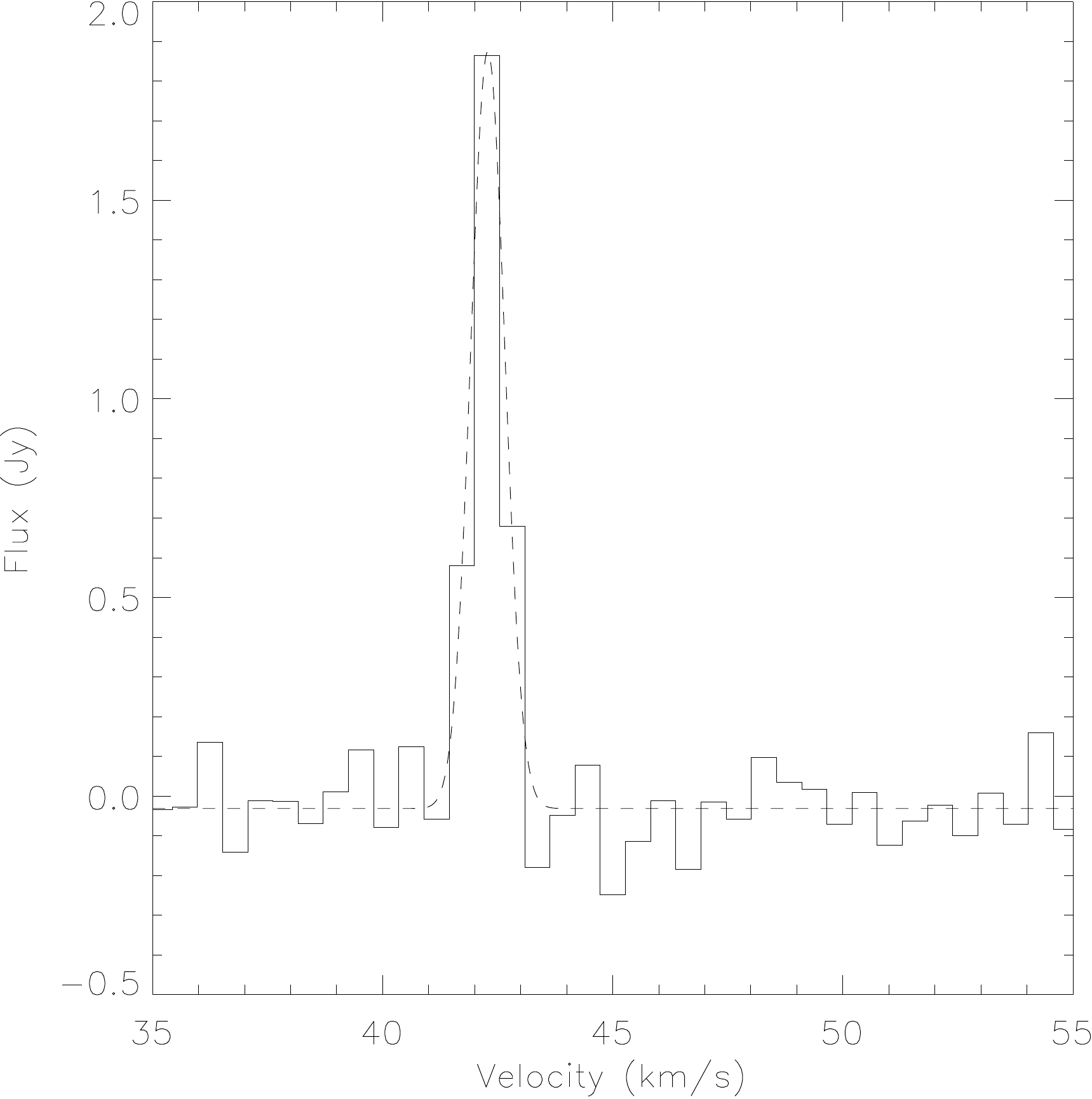}
\includegraphics[width=2.75in]{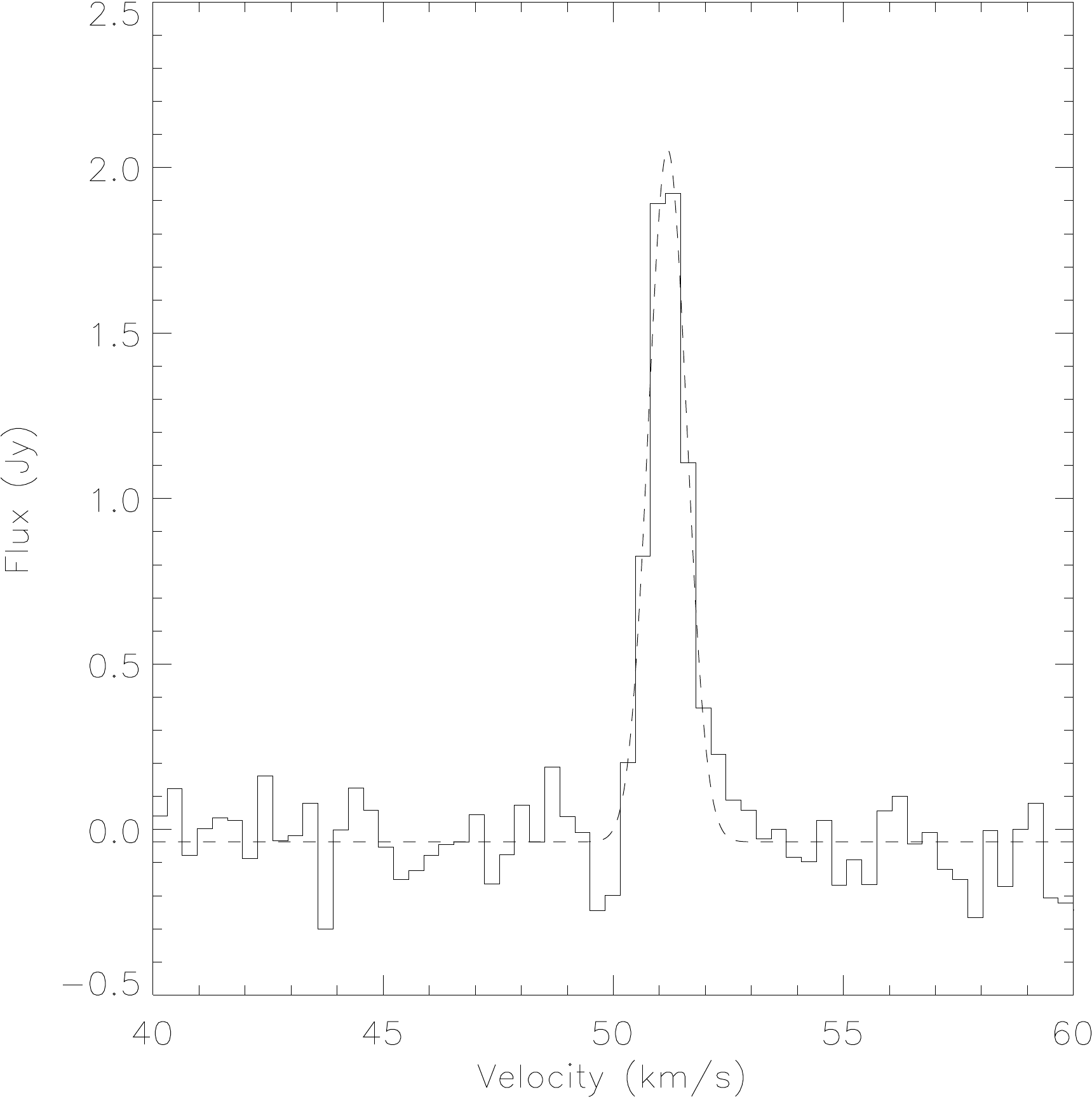}
\includegraphics[width=2.75in]{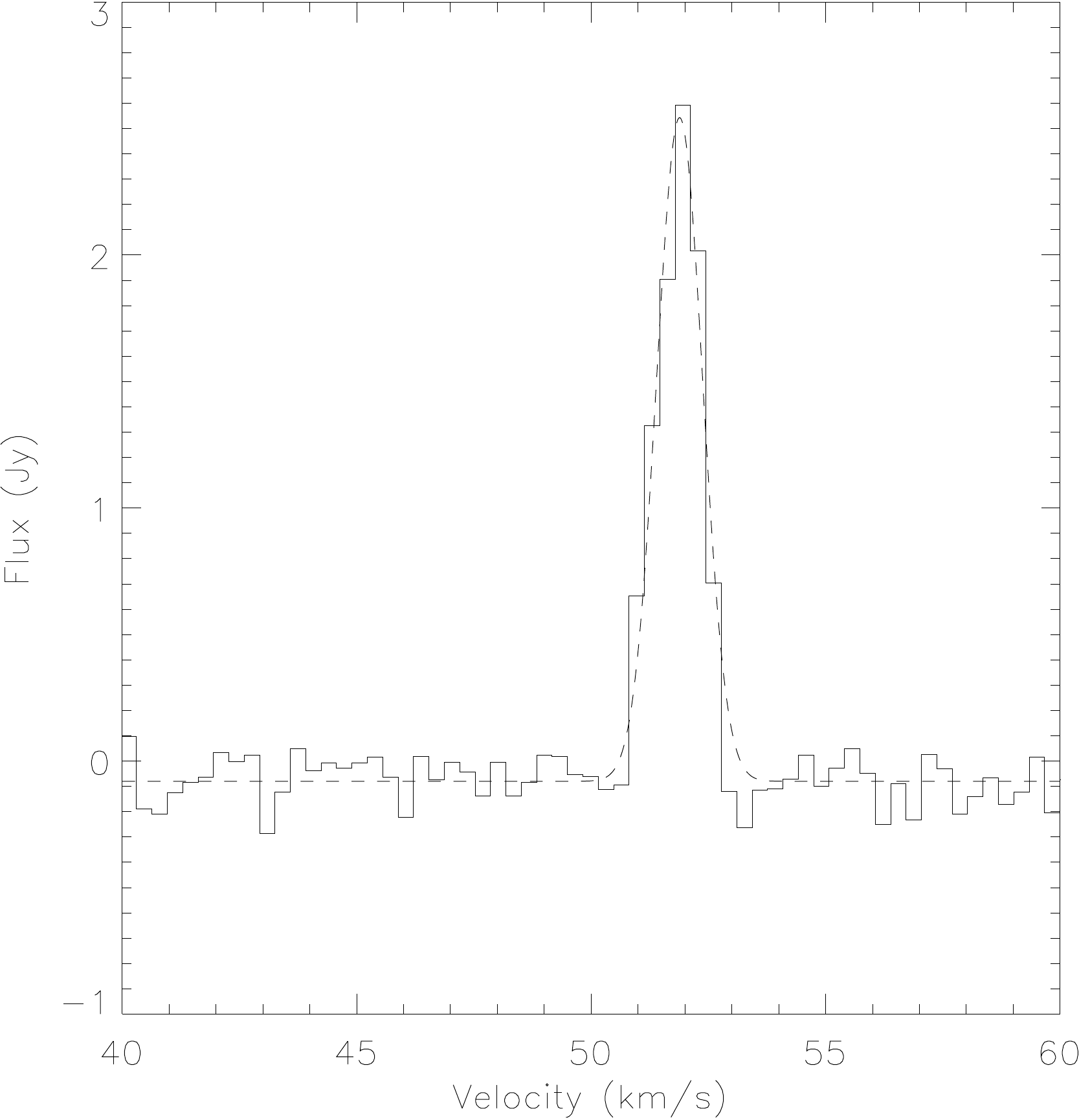}
\includegraphics[width=2.75in]{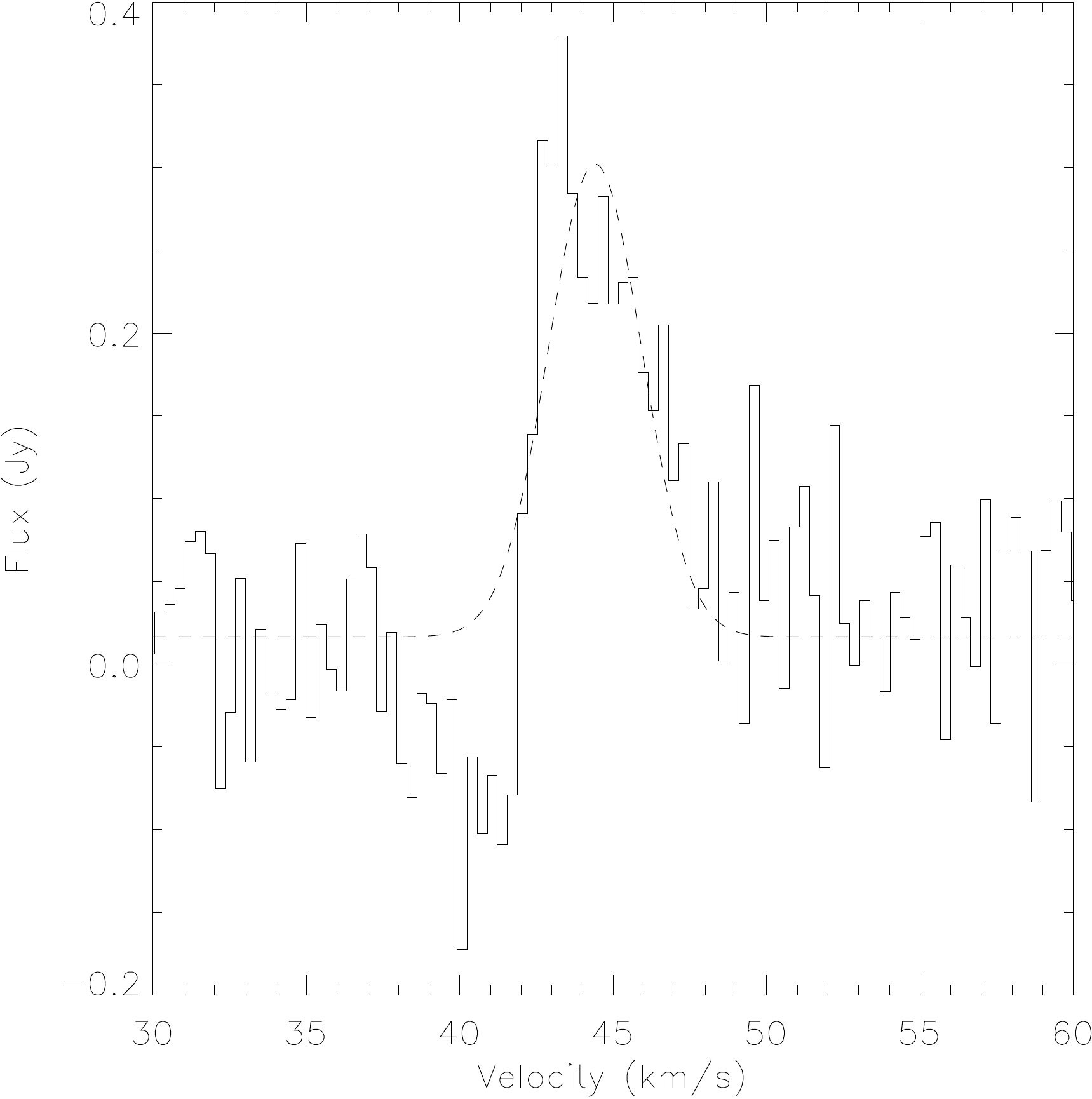}
\end{center}
\caption[Sample Dust Core Spectra] {Spectra of \hco ~at the location
  of several dust cores. Data was integrated with a 7$\arcsec$.5
  tophat kernel. N14-G is at upper left, N22-A upper right, N22-E
  lower left, and N74-C at lower right.
\label{fig-spectra}}
\end{figure}

To find the central velocity of each core, we smoothed the \hco ~data with a 7$\farcs$5-diameter tophat kernel (so as to average over an approximate beamshape) and took the spectrum at the location of the continuum peak. The spectra, which have 0.547 km s$^{-1}$ channels for N14 and 0.328 km s$^{-1}$ channels for N22 and N74, were then fit with Gaussians to find the central velocity. The results are the velocities listed in Table \ref{table-dustcores}, and examples of the spectra are shown in Figure \ref{fig-spectra}.

\section{BUBBLE ENVIRONMENTS \& MORPHOLOGY}
\label{cha:bubbles}

\subsection{N14}
\label{sec-N14}

\begin{figure}
\begin{center}
\includegraphics[width=6.5in]{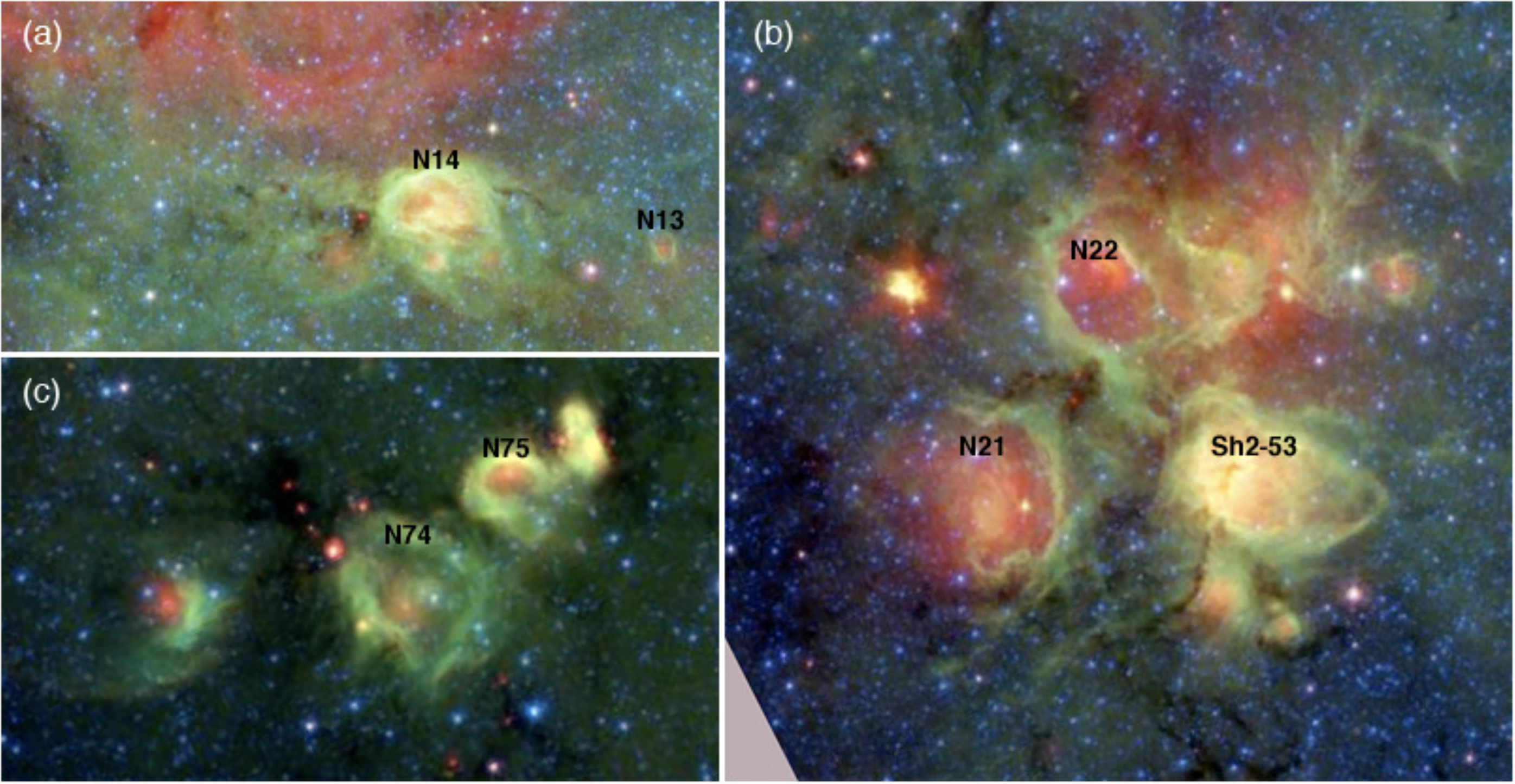}
\end{center}
\caption[The neighborhood of N14, N22, and N74] {Color image of the area
  around the three bubbles. Blue is 3.6 $\micron$ GLIMPSE, green is 8.0 $\micron$
  GLIMPSE, and red is 24 $\micron$ MIPSGAL. Images created from the
  GLIMPSE/MIPSGAL Image Viewer
  (http://www.alienearths.org/glimpse). (a) N14 and its environment. The extended red in the top half of the image is the
  southern half of a large, diffuse bubble. (b) N22 and its environment. The dark nebulosity bordering N22
  to the north is the 65 km s$^{-1}$ cloud. (c) N74 and its environment. The IRDC bordering N74 to the northeast is the location of the
  overdensity of YSOs seen by BW10.
\label{fig-env}}
\end{figure}

The large-scale environment of N14 contains multiple generations of star formation (see Figure \ref{fig-env}a). N14 is part of a filament running east-west, seen in extinction in the infrared against a diffuse 8 $\micron$ background. The filament is obvious to the east and less dramatically to the west. There is a small bubble just to the south of N14.

This whole extended N14-filament region of star formation makes up the southern border of a giant, diffuse bubble, seen most obviously in MIPSGAL and MAGPIS, but also having a noticeable 8 $\micron$ edge on the northern and southern sides. This hierarchical pattern is suggestive though not conclusive of sequential triggered star formation \citep{Oey05}.

The filament includes one prominent IRDC that borders N14 on the east side and contains a protostar seen in 24 $\micron$, which is coincident with dust condensation D-3 (from Table 2 of D10) and dust core J (from this paper).

The \hii region within N14 appears fairly symmetrical, but the PDR is distinctly different on the northern and eastern sides from the southern and western. The northern and eastern sides show a sharply defined rim of intense 8 $\micron$ flux, which is coincident with a prominent filament of \nh ~extending along the PDR. The southern and eastern sides do not have as clearly defined a PDR, but instead have many small filaments of 8 $\micron$ emission, mostly aligned east-west, coincident with the edge of the \hii region. As well, there is a pillar extending into the bubble from the southern side, which is coincident with dust core D.

D10 posit that N14 is expanding into an inhomogenous medium, specifically one that is lower density to the west and is in the process of opening, with the filaments created by ionized gas flowing out. Ionizing radiation acting on a turbulent medium can create filaments and pillars \citep{Grit09} that look much like those seen on the south and west sides of N14, so the molecular cloud on those sides of N14 is likely turbulent. Dust cores A, B, C, D, and E are seemingly associated with the western and southern rims of the bubble. None of them are coincident with YSOs, so they may be filamentary structures in the turbulent cloud aligned along the line of sight. They do not have broader \hco ~line widths or a bigger range of central velocities than the other dust sources around N14, so the turbulence along the filament is limited.

The location of core F is just inside the northern rim of the bubble. The core is co-located with a local peak in 8 $\micron$ flux and a clump of \nh ~emission. This either a turbulent high-density peak exposed by the shock or a higher-density clump on the far side of the bubble that has slowed down the expansion of the shock and is protruding inside the bubble.

The velocities of the dust cores on all sides of the bubble are remarkably similar. In fact, the velocity dispersion within the cores is generally greater than the velocity dispersion among the cores. This suggests that the turbulence of the molecular cloud on scales as large as the bubble ($>$1 parsec) is less than on core-sized scales ($<$0.1 parsec). If BW10 are correct that these bubbles are open rings in flat clouds, it would explain why we do not see cores in the line of sight of the bubble that are blueshifted or redshifted with the swept-up shell. However, if N14 is an open ring, then it would be likely that it would be inclined with respect to our line of sight, and the shell on the side tilted away from us would be redshifted and the side tilted toward us would be blueshifted. As the shock should be moving at speeds of a few km s$^{-1}$, with our velocity resolution we should be able to measure this gradient. We do not see this, which means that the dust fragments are all in pre-existing fluctuations and are not fragmented from a shell swept up by the expanding shock, unless N14 lies coincidentally flat in the plane of the sky.

As regards the \nh emission as shown in Figure \ref{fig-N14N2H}, it is not surprising that the majority of the \nh ~emission is along the bubble rim. There are, however, some clumps of emission towards the bubble interior. If \nh ~would be dissociated inside the bubble as expected, then those clumps must originate in one of three places:

\begin{itemize}
\item
The \nh ~emission could be in a quiescent part of the cloud along the line of sight if the bubble is embedded in a larger molecular cloud.
\item
The expanding shell could have some dense molecular gas on the front and back sides if the bubble is a three-dimensional bubble and not open in the line of sight.
\item
The clumps could be pre-existing density fluctuations and pillars extending inside the bubble.
\end{itemize}

In the third case, then the \nh ~fragments would be correlated with 8-\micron emission, which we do not see. If the second case, then we would expect the fragments in the line of sight of the bubble to be blue-shifted or red-shifted with respect to the ambient cloud, depending on whether they were on the near side or far side of the bubble. We do not see any systematic velocity shift in these fragments, however. There are two likely possibilities: N14 is either (a) still expanding into a molecular cloud in the line of sight, or (b) have pre-existing high-density peaks or turbulent fluctuations that the shock has exposed but left intact. If option (a), the low level of \nh ~emission within the bubble does put limits on the thickness of the cloud. If the molecular cloud the bubble is embedded in were thicker than a few parsecs along the line of sight, the quiescent clouds emitting \nh ~on the far side of the bubble would be closer in number and flux to those along lines of sight neighboring the bubbles.

\subsection{N22}
\label{sec-N22}

Bubble N22 is part of a large star-forming region that includes N21, \hii region Sh2-53 \citep{Sharpless59}, several small bubbles, and many dark filaments and clumps between and around the bubbles that include embedded 24 $\micron$ point sources, likely protostars (see Figure \ref{fig-env}b). Though \citet{Anderson09b} place N22 400 pc more distant than N21 based on their line-of-sight velocities, it seems likely that they are part of the same system and the differences in velocity are most likely due to motions within the cloud and not galactic rotation.

On the west side of N22 (where the CARMA observations were taken) is a blister \hii region with a clear PDR boundary on the south and west and open to the north. In 20 cm and 24 $\micron$, there is no clear boundary between this \hii region and the \hii region within N22, although in 8 \micron there does appear to be a western rim to N22, separating it from this blister region. There might be separate driving sources, or N22 could be open in some direction diagonal to the line of sight and the expanding shock has run into a neighboring cloud. Without detailed velocity information of the ionized gas or identification of the driving sources, the nature of this blister region is ambiguous.

There is plentiful molecular gas in the 51 km s$^{-1}$ component but very little \nh. Our interpretation of this fact is that this velocity component is associated with the blister \hii region east of N22, and that much of the \nh ~has been dissociated.

To the north of N22 and the blister is an area of high IR extinction that is also has significant 1.1 millimeter continuum flux. As we see in the GRS and our \nh ~observations, this is a cloud with velocity $\sim$65 km s$^{-1}$. The \hii region within N22 has a velocity of 50.9 km s$^{-1}$ \citep{Anderson09b}, and the PDR of the western blister is at a similar velocity, as seen in our CARMA observations. There is no flux, either in the GRS or in our observations, at velocities between $\sim$ 58 and 63 km s$^{-1}$. If the 65 km s$^{-1}$ cloud were at the same distance as the blister, the expanding shock should have run into it and led to a sharp northern edge of the \hii region. The fact that its edge is so well-aligned with the border of N22 is highly suggestive, but it seems that the 65 km s$^{-1}$ cloud is in the foreground and the alignment is a coincidence. 

The four dust condensations from D10, as well as our six more northeastern dust cores are associated with the 65 km s$^{-1}$ cloud. Five of our dust cores, as well as condensations D-2, D-3, and D-4, are within a large cluster of protostars seen as point sources at 8 $\micron$ and 24 $\micron$. Our eight southwesterly cores are all in the southern PDR of the blister or the IRDC in the southern foreground. Only D is co-located with a peak of 8 $\micron$ flux (a YSO in the IRDC). Our interpretation is that the rest arise in the cloud behind the PDR.

Cores B and D, which are part of the IRDC, are redshifted with respect to cores A and C by 3 km s$^{-1}$, which is larger than the FWHM of \hco ~seen in these cores. Cores E, F, G, and H are all at the same velocity to within 0.7 km s$^{-1}$.

\subsection{N74}
\label{sec-N74}

The N74 bubble is embedded in a cloud that also contains N75. Several pockets of dark nebulosity are visible around the edges of the two bubbles, and there is a partial bubble to the east that may also be associated with the system (see Figure \ref{fig-env}c). The fluxes at 8 $\micron$, 24 $\micron$, and 20 cm are much lower from N74 than from N14 and N22. 

In 8 $\micron$, N74 has a complete egg-shaped ring of emission that has similar surface brightness on all sides of the bubble. In the center of the bubble is a small area of higher flux, which could be a feature in the PDR on the far side of the bubble.

The \nh ~emission is almost entirely from the dark clouds to the northeast of the bubble and the bubble rims, so does not give much conclusive evidence about N74's three-dimensional structure.

Along the eastern and northeastern sides of N74, BW10 found a significant statistical overdensity of YSOs above the surrounding field, one of only two bubbles around which they found such an overdensity. This coincides with the area mapped by CARMA, and from their Figure 8, appears to include much of the IRDC containing dust cores C and D as well as the eastern bubble rim on which cores A and B reside.

Dust cores A and B are located along the 8 $\micron$ rim. Core B, which has multiple peaks, is co-located with a point source visible at both 24 $\micron$ and 8 $\micron$, so it is likely part of a protostellar system. Dust cores A, B, and D are all in a narrow range of velocities (40.6 - 40.9 km s$^{-1}$). We do not see any evidence of N74 being tilted with respect to the plane of the sky, as no velocity gradient is apparent in our \hco ~data or in GRS data, which has velocity resolution of 0.21 km s$^{-1}$.

Dust cores C and D all lie within the IRDC, $\sim$1.8$\arcmin$ away from the bubble rim. Core C is at $\sim$44.5 km s$^{-1}$, redshifted with respect to the cores closer to the bubble rim. Both have \hco ~linewidths $>$2.8 km s$^{-1}$, much wider than those around the bubble rim.

It is odd that we did not detect more high-mass dust cores, considering the overdensity of YSOs found by BW10. We considered the possibility that N74 is actually at the far kinematic distance of 10.4 kpc, rather than the near distance of 2.8 kpc as previously assumed. At 10.4 kpc, the dust cores we detected would actually range from 590 to 1297 M$_\sun$. If our detection limit was $\sim$600 M$_\sun$ rather than 40 M$_\sun$, it would make sense that we would not detect cores with much higher signal-to-noise than our cutoff, but would imply extremely high mass and high luminosity dust cores and YSOs in the IRDC and bubble rim, which is implausible considering the relatively weak strength of the driving \hii region. If N74 were actually at 10.4 kpc rather than 2.8 kpc, the strength of the \hii region is understated by BW10. Scaling up to the greater distance would indicate that N74 is actually driven by the equivalent of $\sim$0.6 O9.5 stars, which is still significantly weaker than N14 or N22.

\section{FRAGMENTATION AND STAR FORMATION}
\label{cha:fragmentation}

There clearly is star formation occurring on the rims of bubbles, as seen in this paper as well as other recent studies \citep{Zavagno10b}, but it is unclear if the dust cores detected by CARMA are are the progenitors of O stars. Mol 160, an object thought to be an accreting core that will be an O star in the future, has a dust core mass of 500 M$_\sun$ \citep{Mol08}. A few of our cores have masses of that order of magnitude, but are either extended or are unresolved multiple peaks. With the large uncertainties in the flux-mass relation, those objects may deserve further study, as if any will be massive enough to drive their own \hii regions, the triggered star formation would reach the threshold of a positive feedback loop. In N14, the lack of SiO and CH$_3$OH emission indicate that none of the protostars around that bubble are massive enough or developed enough to have driven a shock into their surroundings, as protostellar outflows generally excite SiO and can induce methanol masers.

\citet{DBW07} ran simulations which indicated that an expanding \hii region around an O7 star should sweep up a dense enough cloud of neutral material to fracture into collapsing clumps and form new stars, but the fragmentation occurs once the shell is $\sim$10 pc in radius, an order of magnitude larger than the bubbles we observed. The analytical model that \citet{HI05} present predicts gravitational instabilities to grow in the shell once it reaches a radius of $\sim$3 pc, still significantly larger than N14 or N22. Both studies assume that the medium surrounding the bubble is fairly homogenous.

As we see in our observations, there are large density fluctuations in the clouds surrounding these younger systems. These three objects are not unusual in that regard, as bright dust emission seen in the BGPS and ATLASGAL around bubbles are most often concentrated on one side. Simulations of expanding bubbles in highly inhomogeneous clouds are necessary to reflect the complexity of molecular clouds on parsec scales and investigate the effects this has on the efficiency of triggered star formation.

One important issue is that if the bubbles are open, as BW10 conclude, the triggered star formation efficiency will likely be much lower than expected since a significant fraction of energy will be lost to the champagne flow. The low level of \nh ~seen within N14 and N74 indicate that the molecular cloud through the line of sight is very thin if not entirely open. This is not too surprising, as selection effects of bubbles will preferentially choose objects without strong extinction in front of the \hii region. If they were truly two-dimensional rings, we would expect a velocity gradient due to an angled orientation with respect to the line of sight, but the conclusion that openings in the cloud will lower the efficiency of triggered star formation is relevant. Even setting aside whether or not it is open in the line of sight, N14 appears to also be in the process of opening to the west, so by the time it reaches 3 pc in radius (approximately four times what it is now) it will certainly not be a well-confined spheroid.

An important concern in studying triggered star formation is that correlation does not imply causation. Even if star formation is found to occur in the shells surrounding \hii regions, it is hard to establish that the expanding shock had an important effect since such clouds would likely have active star formation even without the nearby presence of an OB star or cluster. One way of trying to address this concern is to see if the bubble shells are particularly fractured into collapsing clumps over what is seen in a cloud that has not been compressed by the shock.

\begin{figure}
\begin{center}
\includegraphics[width=3.5in]{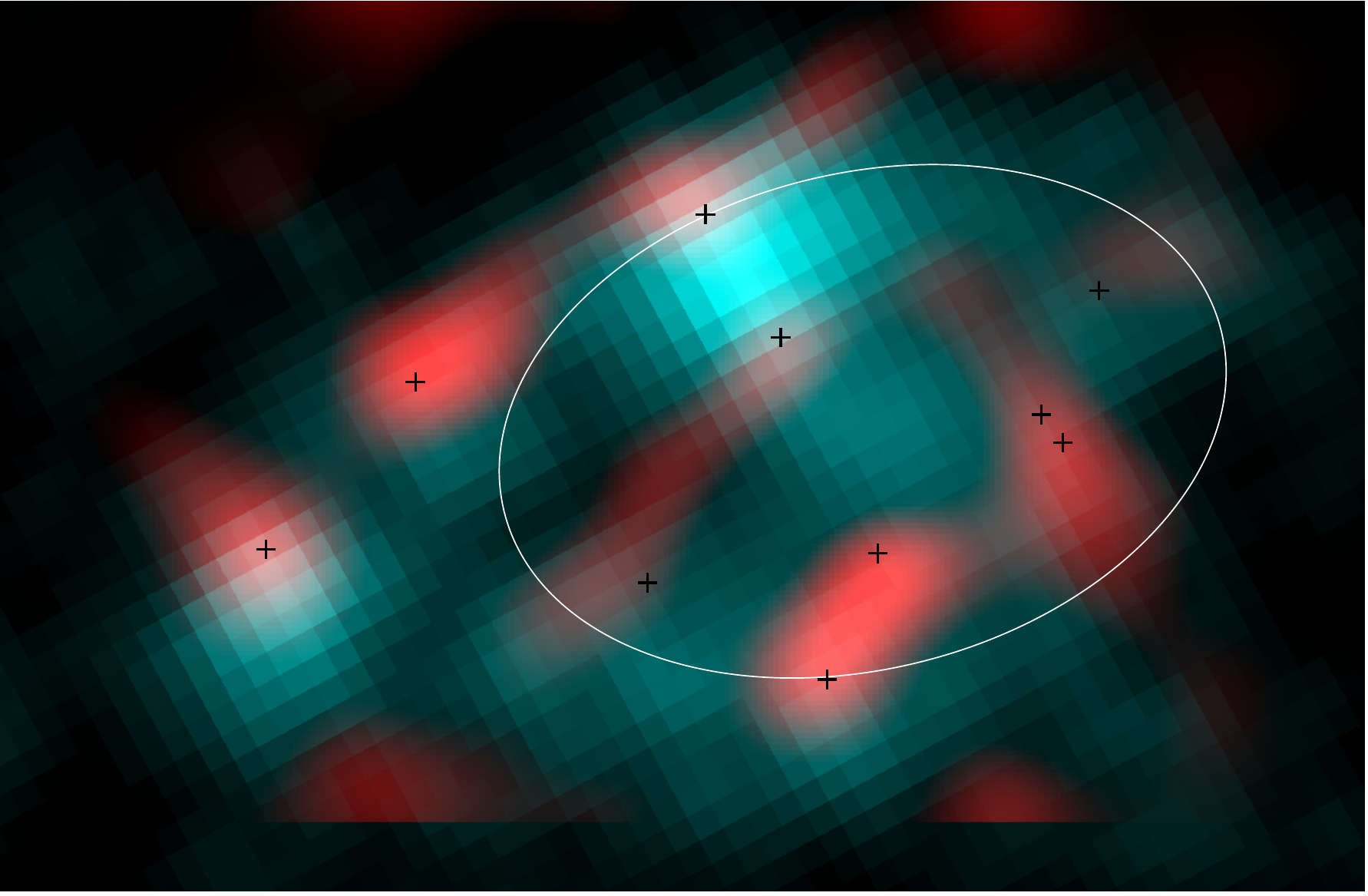}
\includegraphics[width=3.5in]{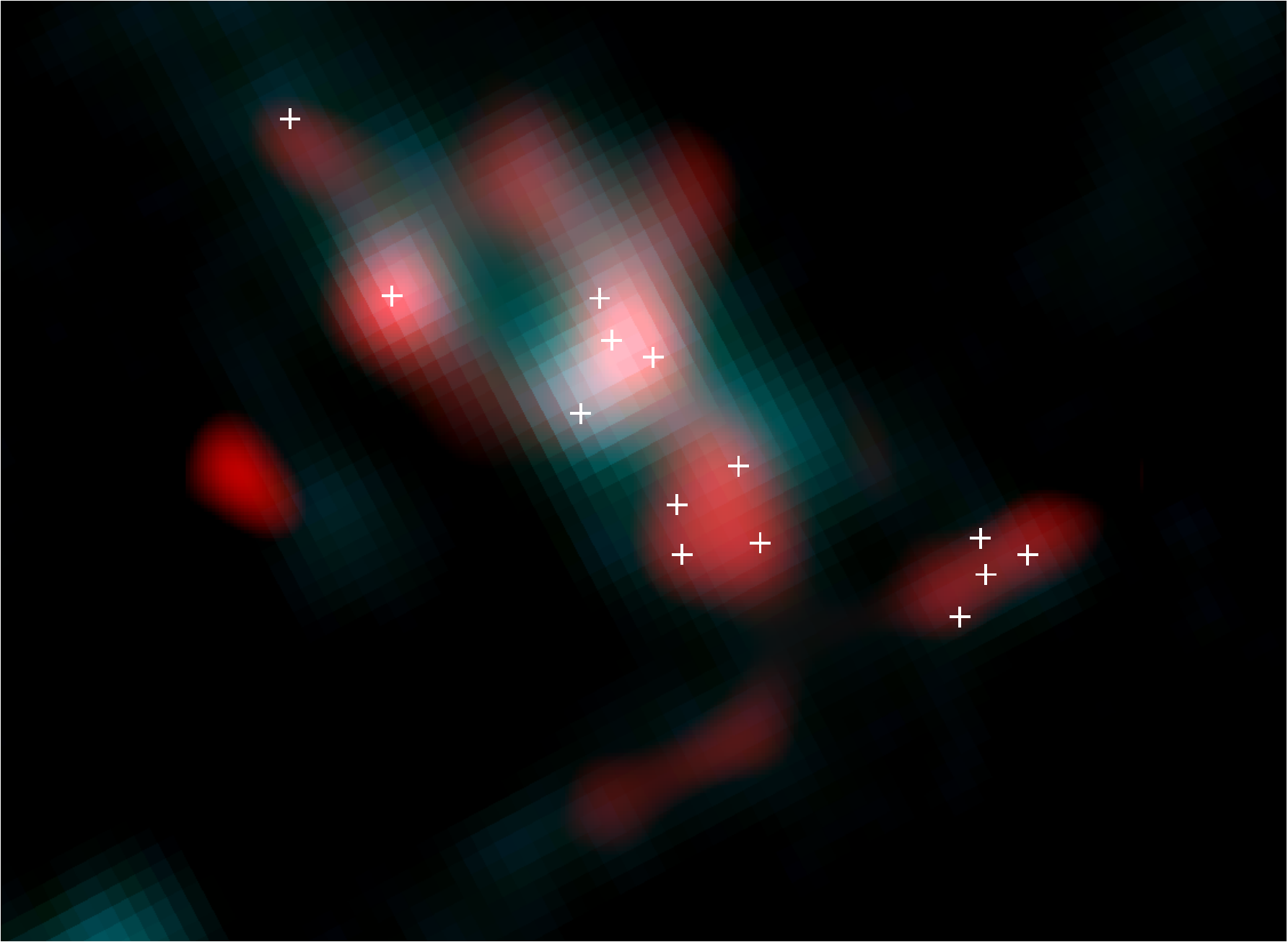}
\includegraphics[width=3.5in]{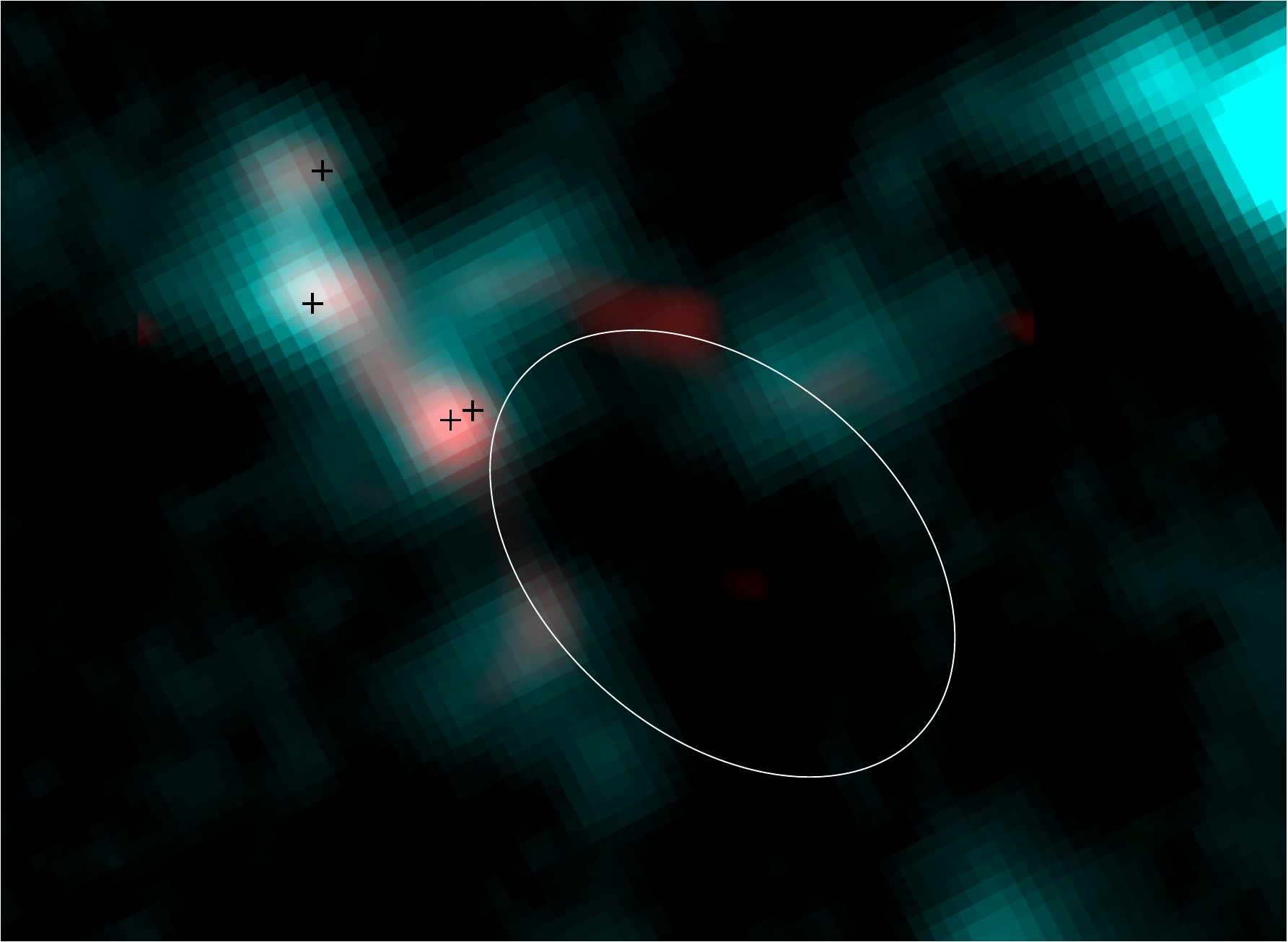}
\end{center}
\caption[Millimeter Dust Color] {Color image of N14 (top), N22 (middle),
  and N74 (bottom). 1.1 mm BGPS data is in turquoise. 3.3 mm
  CARMA data is in red, smoothed with a beam to match the 33$\arcsec$
  resolution of BGPS. Crosses show the locations of the dust
  cores presented in this paper, and white ellipses approximately
  trace the bubble rims of N14 and N74 seen in 8 $\micron$.
\label{fig-dustcolor}}
\end{figure}

To investigate whether the bubbles are inducing fragmentation, we can compare the dust seen by CARMA to the dust emission seen in the BGPS, which includes structures larger than those to which CARMA is sensitive. We took our continuum maps (which were made using a Brigg's visibility weighting of 1 to balance sidelobe suppression but retain sensitivity to as many spatial scales as CARMA could detect) and convolved them with a 33$\arcsec$ beam to match the BGPS data. We then made color images to compare the flux from small-scale structures with the total dust flux. These are shown in Figure \ref{fig-dustcolor}. The BGPS data is in blue and CARMA data in red, so bluer emission has a lower interferometer to single-dish flux ratio, and redder emission has a higher ratio, implying a larger fraction of the flux is in smaller-scale structures.

The ratio of 1.1 mm flux to 3.3 mm flux does depend on things other than the fragmentation. If the dust is at higher temperatures, the ratio will be higher, although this effect is not large. Going from 20 K to 50 K only changes the ratio by 16\%, for instance. If the spectral index of dust emissivity $\beta$ changes drastically, that could skew the ratio, but we will assume a constant $\beta$ and take the ratio as a measure of the fragmentation.

From Figure \ref{fig-dustcolor}, it is apparent that the ratio of fluxes does vary across the maps. The northern part of the N14 bubble is much ``bluer'' than the southern part, so it could be that the higher turbulence along the southern rim caused greater fragmentation. In N22, the area that corresponds to the 65 km s$^{-1}$ cloud, in the northern part of the image, has more blue and white, whereas the clumps in the southern part, which corresponds to the blister PDR, are much redder. Around N74, the clumps along the rim of the bubble, like those near cores A and B, appear slightly redder than the cloud to the northeast, though not drastically so.

This is some evidence that there is greater fragmentation closer to the edges of \hii regions than in dark clouds. To get a better measure of this effect we would need interferometric data of more bubbles, or at higher signal-to-noise, but we present this analysis as a concept worth further investigation. Single-dish observations of optically thin molecular lines, like C$^{18}$O, combined with CARMA observations, would give more robust results, as the signal-to-noise of C$^{18}$O is higher than the 3.3 mm dust and would not be affected by uncertainties in dust emissivity. CO and its isotopes are often depleted onto grains in dense dust cores, but such analysis could still reveal fragmentation on clump-sized scales, and velocity information could be used to more easily separate components like the two clouds seen in N22. Reanalyzing dust continuum using the soon-to-be-public ATLASGAL data, which is at higher-resolution than BGPS, could also add weight to this result.

\section{CONCLUSIONS}
\label{cha:conclude}

We present exploratory observations of three infrared bubbles (N14, N22, and N74) thought to be expanding \hii regions around young massive stars. CARMA has resolved structure in molecular lines of \nh, \hco, HCN, CS, C$^{18}$O, $^{13}$CO, and 3.3 mm continuum down to the 5$\arcsec$ resolution of the observations, which corresponds to sizes of $\sim$0.1 pc.

\nh ~emission, which traces the gas in dark clouds not exposed to the \hii region, reveals that N14 is expanding into a very inhomogeneous cloud, with much greater density on the northern and eastern borders of the bubble. It also suggests that N14 has not entirely blown an open hole through its parent molecular cloud, though it is in the process of opening. The \nh gas around N22, meanwhile, is seen in two distinct velocity components, one at the bubble velocity of 51 km s$^{-1}$ and one at 65 km s$^{-1}$. Much of the dust and star formation seen neighboring N22 is in the 65 km s$^{-1}$ cloud, which appears to be independent from the bubble shell.

We have found 28 compact 3-mm continuum sources around the three bubbles. Only six of the 28 are associated with 24 $\micron$ point sources seen in MIPSGAL. We have measured their fluxes and estimated their masses, which vary from 36 to 733 solar masses. We found CS was depleted in dust cores, as CS emission was only detected from five of the 28 continuum sources. \nh ~was often found near to dust cores but tended to be depleted at very small radii. \hco ~and HCN were the best correlated molecular lines with dust cores, with strong emission found at the location of nearly all continuum sources. Analyzing the spectra of \hco ~and HCN molecular lines gave us the velocities of the dust cores, confirming their associations with the bubbles and surrounding clouds.

Our results are consistent with the bubbles being open towards our line of sight, but they may not be entirely open rings as \citet{BW10} conclude. We see some signs that the far side of N14 and N74 are still bordering molecular clouds, and do not see any velocity gradient across them, as would be likely for two-dimensional objects.

By comparing single-dish and interferometric observations, the latter of which will resolve out extended structures, we find evidence for some \hii region shells being fragmented more than nearby molecular clouds. This adds credibility to the idea of dense shells around \hii regions being sites of triggered star formation, although our bubble morphologies, with significant turbulence and cloud inhomogeneities, suggest that current models are too simplistic.

\acknowledgments{}
I must thank my graduate adviser Al Harper profusely for his wisdom and advise throughout this project. For this paper I am grateful to Claudia Cyganowski, Dan Marrone, Grace Wolf-Chase, and Ed Churchwell for many valuable conversations and ideas. I also want to acknowledge the staff at OVRO for their support during my trips to CARMA.

Support for CARMA construction was derived from the Gordon and Betty Moore Foundation, the Kenneth T. and Eileen L. Norris Foundation, the James S. McDonnell Foundation, the Associates of the California Institute of Technology, the University of Chicago, the states of California, Illinois, and Maryland, and the National Science Foundation. Ongoing CARMA development and operations are supported by the National Science Foundation under a cooperative agreement, and by the CARMA partner universities.

This work is based in part on observations made with the Spitzer Space Telescope and has made use of the NASA/IPAC Infrared Science Archive, both of which are operated by the Jet Propulsion Laboratory, California Institute of Technology under a contract with NASA.

\bibliography{Bibliography}
\bibliographystyle{apj}

\end{document}